\documentclass{vgtc}                          

\graphicspath{{figures/}{pictures/}{images/}{./}} 

\usepackage{times}                     
\usepackage{comment}

\usepackage{tabu}                      
\usepackage{booktabs}                  
\usepackage{lipsum}                    
\usepackage{mwe}                       
\usepackage{enumitem}

\setlist{nosep}

\usepackage{mathptmx}                  
\usepackage{amssymb}
\usepackage{amsmath}

\DeclareMathAlphabet{\mathcal}{OMS}{cmsy}{m}{n}
\DeclareMathOperator{\bin}{bin}

\usepackage{subfig}

\newcommand{\newtext}[1]{\textcolor{black}{#1}\xspace}
\newcommand{\myparagraph}[1]{\vspace{1mm} \noindent \textbf{#1}}
\newcommand{\ie}{i.e.,\xspace}
\newcommand{\etal}{et al.\xspace}

\onlineid{1022}

\vgtccategory{Application}

\vgtcinsertpkg

\title{Topology-Aware Volume Fusion for Spectral Computed Tomography via Histograms and Extremum Graph}

\author{Mohit Sharma\thanks{e-mail: mohit.sharma@liu.se} %
\and Emma Nilsson\thanks{e-mail: emma.nilsson@liu.se}
\and Martin Falk\thanks{e-mail: martin.falk@liu.se}
\and Talha Bin Masood\thanks{e-mail: talha.bin.masood@liu.se}
\and Lee Jollans\thanks{e-mail: lee.jollans@liu.se}
\and Anders Persson\thanks{e-mail: anders.persson@liu.se}
\and Tino Ebbers\thanks{e-mail: tino.ebbers@liu.se}
\and Ingrid Hotz\thanks{e-mail: ingrid.hotz@liu.se}}
\affiliation{\scriptsize Link\"oping University, Sweden}
\teaser{
  \centering
  \includegraphics[width=\linewidth]{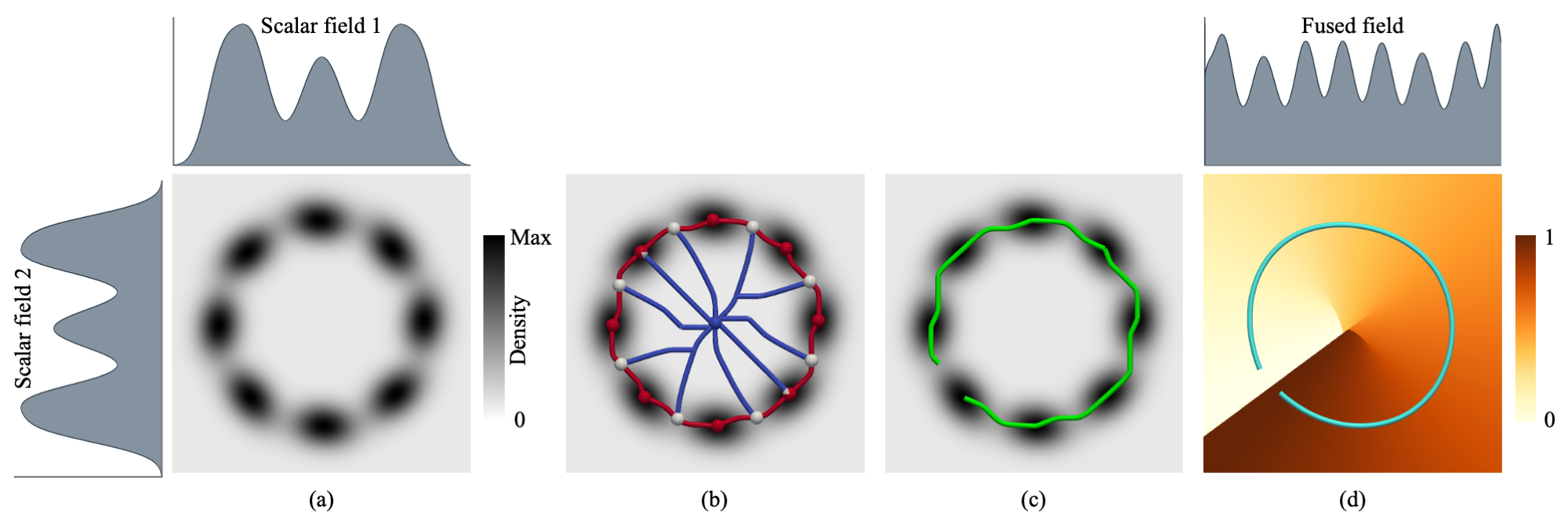}
  \caption{Volume fusion of two hypothetical volumes producing a 2D histogram resulting in eight Gaussian-distributed peaks on a circular path. (a) Synthetic 2D histogram; x-axis: scalar field from volume 1, y-axis: from volume 2; projected 1D histograms of individual volumes shown along the axes are unable to separate all features. (b) Morse–Smale complex of histogram density field; red: maxima, white: saddles, blue: minima. (c) Minimum spanning tree (green) of extremum graph, \ie restricted to maxima and saddles; the longest simple path of the minimum spanning tree captures major ridge structure, traversing all eight Gaussian peaks. (d) Smooth spline (aqua) fitted to the path extracted in (c); fused scalar field generated by projecting histogram points to this spline; 1D histogram reveals the eight key features.}
  \label{fig:teaser}
}

\abstract{
    Photon-Counting Computed Tomography (PCCT) is a novel imaging modality that simultaneously acquires volumetric data at multiple X-ray energy levels, generating separate volumes that capture energy-dependent attenuation properties. Attenuation refers to the reduction in X-ray intensity as it passes through different tissues or materials, which depends on their density and atomic composition. This spectral information enhances tissue and material differentiation, enabling more accurate diagnosis and analysis. However, the resulting multivolume datasets are often complex and redundant, making visualization and interpretation challenging. To address these challenges, we propose a method for fusing spectral PCCT data into a single representative volume that enables direct volume rendering and segmentation by leveraging both shared and complementary information across different channels. Our approach starts by computing 2D histograms between pairs of volumes to identify those that exhibit prominent structural features. These histograms reveal relationships and variations that may be difficult to discern from individual volumes alone.
    Next, we construct an extremum graph from the 2D histogram of two minimally correlated yet complementary volumes—selected to capture both shared and distinct features—thereby maximizing the information content. The graph captures the topological distribution of histogram extrema. By extracting prominent structure within this graph and projecting each grid point in histogram space onto it, we reduce the dimensionality to one, producing a unified volume. This representative volume retains key structural and material characteristics from the original spectral data while significantly reducing the analysis scope from multiple volumes to one. The result is a topology-aware, information-rich fusion of multi-energy CT datasets that facilitates more effective visualization and segmentation.
} 


\begin{document}


\firstsection{Introduction}
\label{sec:intro}
\maketitle
Computed Tomography (CT) is one of the primary imaging modalities used daily in clinical practice. CT scanners produce volumetric data expressed in Hounsfield units (HU), a standardized, dimensionless scale that quantifies X-ray attenuation coefficients. These values reflect the density of materials or tissues.
Experienced practitioners can efficiently define transfer functions\newtext{, which map scalar field values to visual properties like color and opacity,} for volume rendering to explore 3D visualizations of the human body. However, conventional CT scans struggle to distinguish between tissues with similar attenuation values, particularly among soft tissues, limiting diagnostic capabilities in some cases.
Photon-Counting Computed Tomography (PCCT) is an emerging imaging technology, with the first clinical systems now in operation, aimed at overcoming the limitations of conventional CT. Unlike conventional CT, which measures the cumulative X-ray energy, PCCT counts and converts individual X-ray photons directly into electrical signals. This has the potential to significantly improve image quality~\cite{Flohr:2023:PCCD}. Expected benefits include higher spatial resolution, electronic noise elimination, improved contrast-to-noise ratio, lower radiation doses, and intrinsic spectral (multi-energy) imaging, which enables better tissue differentiation and material characterization~\cite{Greffier2025}.

However, in practice, we are still far from fully exploiting the potential of this new technology. The acquired imaging data is rapidly increasing in both size and complexity, presenting significant challenges in terms of data handling, accessibility, and for the standard data analysis pipelines.
One particular challenge that we are approaching with this work is the efficient use of the energy-specific photon signals, which is a set of co-registered volumes, each corresponding to a different energy range, thereby encoding energy-dependent attenuation properties of tissues and materials. Although for standard CT data, well-established data analysis pipelines including automatic image segmentation and transfer function generation are available, they cannot be applied directly to the new data.

While several general-purpose methods have been proposed for designing multi-dimensional transfer functions, such as parallel coordinates, multiple linked views, and dimensionality reduction techniques, these approaches are often incompatible 
with standard CT image analysis workflows and don’t exploit the specific properties of PCCT data.  The multi-dimensional nature of PCCT data demands the development of new methods tailored to its unique characteristics.

The goal of this work is to enable the integration of traditional CT image analysis methods and pipelines within the context of PCCT data. The core idea is to generate a fused volume that takes advantage of the unique properties of multichannel data, specifically, the presence of both complementary\newtext{---features that are distinctly visible in only one of the volumes---and} overlapping information\newtext{---features that appear in both volumes but may differ in intensity or contrast.}

\paragraph{Contributions}
    \begin{itemize}
        \item A novel extremum graph-based technique to compute a fused volume from multi-energy volume data.
        \item Dimensionality reduction that retains important structure and is topology-aware.
        \item Enhanced rendering and segmentation using the fused volume.
    \end{itemize}
The effectiveness of the method is demonstrated using a synthetic dataset and two real PCCT datasets.


\begin{figure*}[!t]
    \centering
   \includegraphics[width=\textwidth]{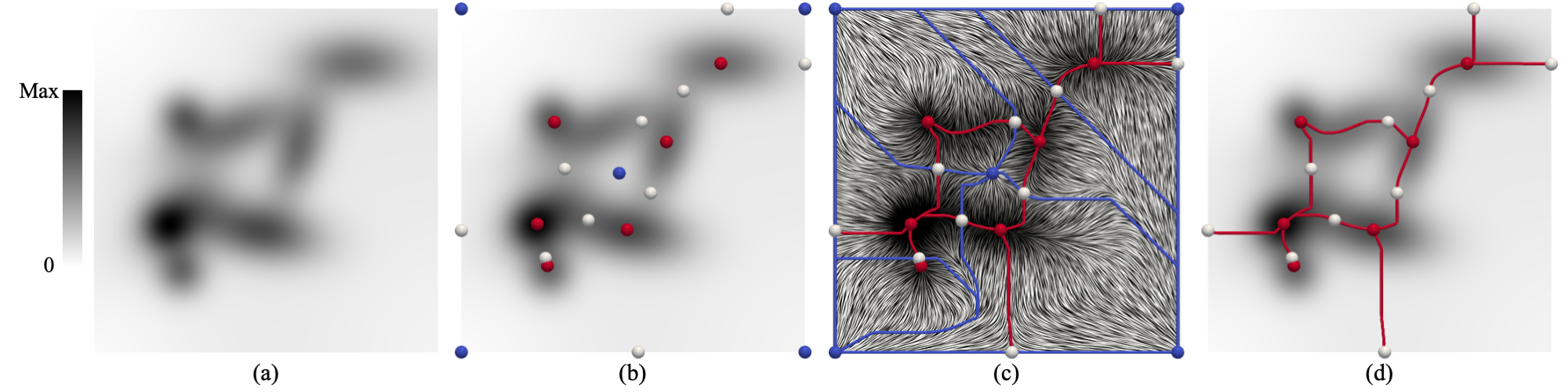}

    \caption{2D scalar field topology. (a) A scalar field $f:\mathcal{M}\rightarrow\mathbb{R}$ defined over a 2D manifold. (b) Critical points ($\nabla f(p) = 0$): minima (blue), maxima (red), and saddles (white). (c) Morse–Smale complex, segmenting $\mathcal{M}$ into regions of uniform gradient behavior. Blue and red lines represent the descending and ascending separatrices, respectively, which separate regions whose integral lines originate from the same minimum $m^-$(blue) or converge to the same maximum $m^+$(red). (d) Extremum graph capturing connectivity between maxima and saddles using the ascending separatrices. A similar graph can be extracted to capture the minima-saddle connectivity using descending separatrices.}
    \label{fig:sft-background}
\end{figure*}

\section{Related work}
\label{sec:relatedWork}
This paper focuses on generating a single fused volume from a multivolume dataset using topological data analysis (TDA), with a particular emphasis on extremum graphs. This section reviews prior work on volume rendering of multivolume data, as well as relevant techniques in TDA and extremum graph-based analysis. \newtext{We use the terms \emph{multivolume}, \emph{multichannel}, and \emph{multivariate} data interchangeably depending on context. 
\emph{Multivolume} denotes multiple co-registered volumes acquired under different parameters (e.g., X-ray energies in PCCT). 
\emph{Multichannel} denotes treating these volumes as channels of a single dataset, analogous to spectral bands in hyperspectral imaging. 
\emph{Multivariate} is the general term for data with multiple attributes per spatial location, regardless of origin.
}

\subsection*{PCCT and volume rendering of multi-channel data}
Photon-counting computed tomography (PCCT) is a relatively recent imaging modality, and much of the current literature focuses on the fundamental aspects of the technique, such as system characterization and image reconstruction algorithms~\cite{Bie2023}. Consequently, there is still limited research addressing visualization and segmentation techniques fully leveraging the spectral data. A recent review by Alves et al.~\cite{Carolina-Alves2024} highlights several critical challenges in this domain, including the lack of publicly available PCCT datasets and the inherent complexity of 3D spectral data, which poses significant challenges in deep learning workflows due to computational limitations. Some work has explored the use of Convolutional \newtext{DenseNets} for multi-material decomposition and classification in PCCT~\cite{Wu2019}, which could potentially impact transfer function design by leveraging such classification outputs.
Beyond PCCT-specific applications, there have been various efforts to adapt transfer function design to spectral imaging contexts. For instance, Noordmans et al.~\cite{Noordmans2000} proposed an early spectral extension of the volume rendering integral. Strengert et al.~\cite{Strengert2006} built on that work, demonstrating GPU-based raycasting methods that simulate spectral effects such as selective absorption and dispersion using physically-based optical properties. More recently, Falk et al.~\cite{Falk2017} introduced a transfer function design toolbox tailored for full-color volumetric datasets utilizing perception-based colorspaces. In astrophysics, Alghamdi et al.~\cite{Alghamdi2023} presented a spectral volume rendering approach that incorporates wavelength-dependent emission–absorption modeling to visualize Doppler effects.

Multichannel or spectral data generally fall under the category of multivariate data. A considerable amount of research has been devoted to developing interfaces for designing multidimensional transfer functions (TFs) to visualize such data. \newtext{Kniss~\etal~\cite{Kniss2002} introduced interactive 2D transfer functions for volume rendering and laid the groundwork for many subsequent approaches.} These interfaces typically rely on coordinated, linked views, often supported by clustering, segmentation, or dimensionality reduction techniques applied to the high-dimensional attribute space.
Dobrev et al.~\cite{Dobrev2011} utilize a clustering hierarchy to enable interactive TF generation. Jankowai et al.~\cite{Jankowai2020} present an interface that leverages cluster representatives to design TFs for rendering tensor fields. Wang et al.~\cite{Wang2012} automatically generate TFs using a Morse decomposition of a 2D density plot. Similarly, Cai et al.~\cite{Cai2017} propose an approach that combines topology-preserving dimensionality reduction with a subsequent clustering step. Zhao and Kaufman~\cite{Zhao2010} integrate local linear embedding with a parallel coordinates interface. Liu et al.\cite{Liu2014} assume that the high-dimensional attribute space can be effectively represented through a set of linear projections and use dynamic, animated projections to guide users through the space. More recently, Huang et al.~\cite{Huang2025_bimodal} introduced a method for bimodal visualization in the context of advanced manufacturing. Their approach creates a combined view of co-registered bimodal data by applying automatic topological segmentation to a bivariate histogram, which is then used as an interface for interactive data exploration.
In addition to these techniques, some methods, similar to ours, aim to fuse the multivariate data into a single volume that can be rendered using standard techniques. Haidacher et al.~\cite{Haidacher2008} perform data fusion based on information-theoretic measures. Kim et al.~\cite{Kim2011} apply Isomap\newtext{,} local linear embedding, \newtext{and} principal component analysis for dimensionality reduction. 
For a more comprehensive overview, we refer the reader to the state-of-the-art report on transfer functions by Ljung et al.~\cite{Ljung2016}.

\subsection*{TDA and extremum graphs}

Topological Data Analysis~(TDA) offers a robust framework for characterizing the shape and structure of complex datasets. It is particularly effective for feature-based analysis of scalar fields, i.e., real-valued functions defined over spatial domains. The topological methods used in scalar field analysis can be broadly categorized into two principal approaches: level set-based methods (such as merge trees, contour trees, and Reeb graphs) and gradient-based methods (including the Morse complex, Morse-Smale complex, and extremum graphs)~\cite{YanScalar2021}.
These topological structures are well-suited for hierarchical representations of scalar data, enabling the extraction of features across multiple scales~\cite{HeineScalar2016}. 

The Morse complex partitions the domain of a scalar field according to its critical points and the associated gradient flow, resulting in a segmentation into regions of uniform gradient behavior. 
Correa et al.~\cite{Correa2011_TopoSpines} introduced \emph{extremum graph}, a subset of Morse complex, that still captures the key features of the scalar field without the complexity of the complete Morse or Morse-Smale complex. The extremum graph emphasizes the connectivity of extrema (minima or maxima) via saddle points, offering a compact representation of the scalar field's most salient topological characteristics. Using topological persistence-based filtration~\cite{edelsbrunner2002topological}, insignificant features -- often corresponding to noise -- can be filtered out, while preserving meaningful structures such as ridges and valleys. Efficient GPU-based algorithms have recently been developed for the computation of extremum graphs \cite{Ande2023_tachyon}. Morse-Smale complex and extremum graphs have been used for analysis of data originating from a wide range of application domains, including material science \cite{Pandey2021_Granular}, chemistry \cite{Gyulassy2016_ion_battery}, meteorology~\cite{engelke2021topology, nilsson2022exploring} and cosmology \cite{sousbie2011persistent,Shivashankar2016_Felix}.

\section{Background}
\label{sec:background}
This section introduces the necessary scalar field topology background required to understand the rest of the paper, with a focus on the extremum graph structure that is central to the proposed method. Morse theory offers a robust theoretical framework for extending topological structures to n-dimensional domains; however, the following discussion is restricted to two dimensions, as the method deals with extremum graphs derived from 2D histograms.

\myparagraph{Critical points and Morse functions.}
Consider a continuous and differentiable scalar function $f: \mathcal{M} \rightarrow \mathbb{R}$, where $\mathcal{M}$ is a 2D manifold (see \autoref{fig:sft-background}(a)). A point $p \in \mathcal{M}$ is called a critical point if the gradient vanishes at that point, \ie $\nabla f(p) = 0$. Critical points are classified as minima, maxima, or saddles by analyzing the signs of the eigenvalues of the Hessian matrix at that point. In \autoref{fig:sft-background}(b), blue, red, and white indicate minima, maxima, and saddles, respectively. A critical point is called nondegenerate if the Hessian at that point is invertible. A function $f$ is a Morse function if all critical points are nondegenerate and have distinct scalar values. Any continuous and differentiable scalar function can be perturbed by an infinitesimally small amount to satisfy these conditions.

\myparagraph{Morse-Smale complex.} The Morse-Smale complex is a topological structure that segments $\mathcal{M}$ into regions having uniform gradient behavior of the scalar field $f$. Consider a curve $c \in \mathcal{M}$ whose tangent at every point $p$ is aligned with the gradient $\nabla f$; such curves are called integral lines. In \autoref{fig:sft-background}(c), integral lines are visualized using the line integral convolution (LIC) method. The set of integral lines that originate from a minimum $m^-$ defines the ascending manifold $\mathcal{A}(m^-)$, while those that terminate at a maximum $m^+$ define the descending manifold $\mathcal{D}(m^+)$. These manifolds segment $\mathcal{M}$, and their boundaries are known as the ascending and descending separatrices, respectively. In \autoref{fig:sft-background}(c), these separatrices are shown as blue and red line segments. The intersection of the ascending and descending manifolds yields the Morse-Smale segmentation of the domain, in which each cell consists of all integral lines that originate at the same minimum $m^-$ and terminate at the same maximum $m^+$.

\myparagraph{Extremum graph.} A Morse-Smale complex consists of two extremum graphs, as introduced by Correa~\etal~\cite{Correa2011_TopoSpines}. An extremum graph is a substructure of the Morse–Smale complex that captures the connectivity between critical points: between maxima and saddles (maximum graph) or between minima and saddles (minimum graph). The structure preserves ridge and valley structures in the scalar field. Additionally, a persistence-based simplification~\cite{edelsbrunner2002topological} helps remove saddles that connect to the same maximum (in the maximum graph) or the same minimum (in the minimum graph), resulting in a simplified representation of the underlying scalar field. \autoref{fig:sft-background}(d) shows the maximum graph of the 2D scalar field $f$, where maxima and saddles are connected via ascending separatrices.
In this paper, we focus exclusively on the maximum graph, and the term ``extremum graph'' refers to the maximum graph unless stated otherwise. 

\myparagraph{Discrete setting.} To represent real data and enable computation, $\mathcal{M}$ is modeled as a simplicial complex. In 2D, this corresponds to a $2$-complex consisting of vertices~(points), edges~(line segments), and faces~(triangles). The scalar function $f$ is defined at the vertices and interpolated linearly over edges and triangles, resulting in a piecewise linear scalar field. To ensure that $f$ behaves as a Morse function, \ie all critical points are nondegenerate and have distinct values, simulation of simplicity~\cite{edelsbrunner_simulation_1990} is employed to symbolically eliminate degeneracies. 

\begin{figure*}[!t]
    \centering
   \includegraphics[width=\textwidth]{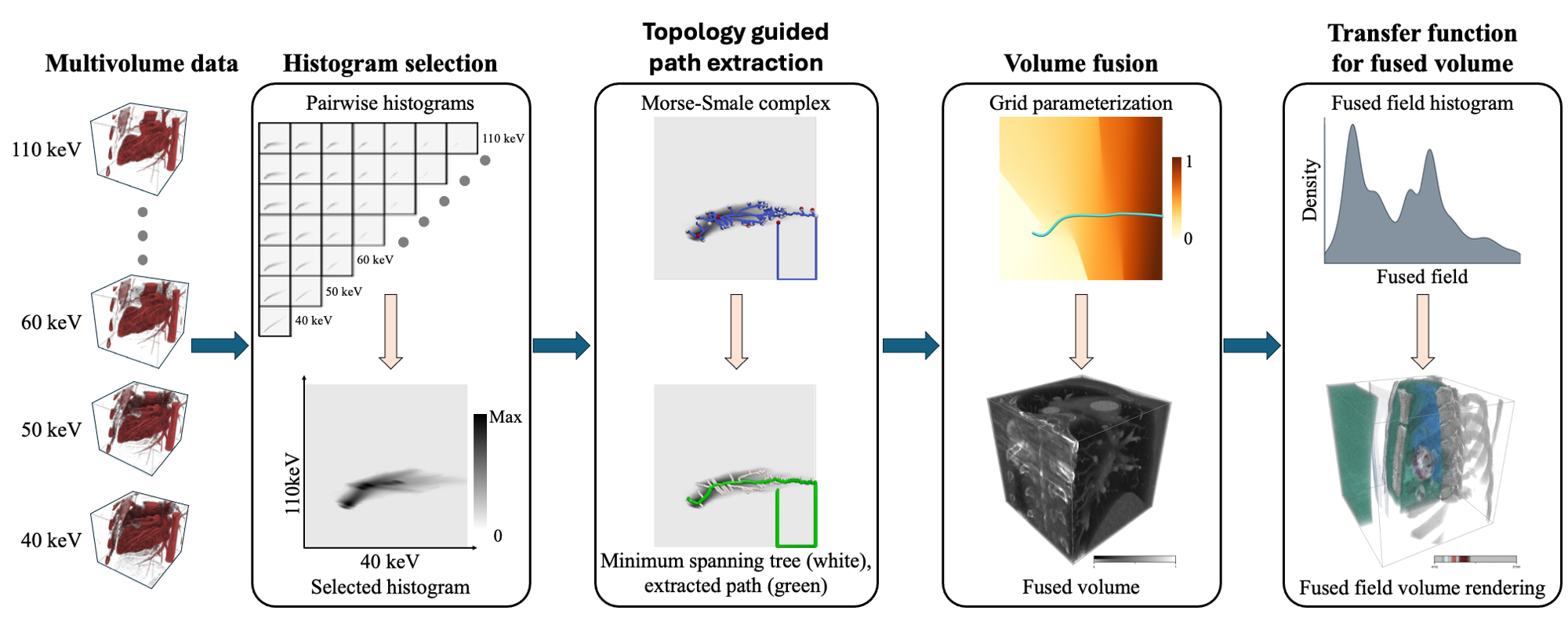}
    \caption{Overview of the fusion pipeline. 
      Multivolume data is taken as input to compute pairwise histograms.
      The histogram capturing the most prominent features is selected. 
      A Morse–Smale complex is computed on this histogram to extract the extremum graph of maxima and saddles. 
      The longest path in this graph is computed and smoothed via spline fitting. 
      Histogram grid points are then projected onto the spline to generate a fused scalar field. 
      The fused volume preserves prominent joint features and is used for volume rendering and transfer function design.} 
    \label{fig:pipeline}
\end{figure*}

\section{Method}
\label{sec:method}
\autoref{fig:pipeline} provides an overview of the method pipeline. Multivolume data is taken as input, and pairwise histograms are computed. One histogram is selected to construct the Morse–Smale complex. From this complex, a meaningful path is extracted to capture the `main ridge' of the data. The histogram is mapped onto this path and parameterized accordingly, enabling the generation of a fused volume based on this parameterization. This fused volume serves as the basis for both rendering and analysis. The subsequent subsections describe each step of the pipeline in more detail using the 2D scalar field in \autoref{fig:sft-background}(a) as a running example.

\subsection{Histogram-based volume pair selection}
\label{sec:histselection}

Given a multispectral PCCT dataset, \newtext{where each volume encodes CT values as a scalar field over $\mathbb{R}^3$ corresponding to a specific energy level, }we begin with computing pairwise 2D histograms for all energy levels and the corresponding correlation between them.
The pairwise Pearson correlation provides a rough indication which of the pairs are of potential interest in addition to the histograms.
As the volumes of a PCCT dataset are co-registered and feature the same dimensions, a 2D histogram of size $n \times n$ for two input volumes $V_1$ and $V_2$ is obtained by binning and counting the pairwise values $(V_1(t), V_2(t))$ of all voxels $t$ in the input volume.
Since the different energy levels are strongly correlated, this typically results in a low number of small peaks (see Figure~\ref{fig:heartpipeline}).
To make less prominent features and low-density regions visible, we therefore apply a logarithmic normalization to each $\bin(i,j)$ of the 2D histogram with respect to the maximum bin count
\begin{equation}
    \bin'(i,j) = \log\left(\bin(i,j) + 1\right) / \log\left(\max_{x,y \in [n]\times[n]} \bin(x,y) + 1\right).
\end{equation}
The resulting histograms mirror the correlation coefficients.
A low correlation indicates a more spread out, and therefore more interesting, histogram capturing the most features, whereas highly correlated volumes yield a linear diagonal histogram.



\subsection{Topology guided path extraction}
\label{sec:pathExtraction}
\newtext{In a 2D histogram of two volumes, critical points correspond to regions of high (maxima) or low (minima) joint occurrence of scalar values, while saddle points indicate transitions between such regions. High occurrence or density typically corresponds to a specific material. Analyzing the topology of these points allows us to trace ridges that correspond to different materials or tissues across both volumes~\cite{Ljung2016TF}.} We compute the Morse--Smale (MS) complex of the selected 2D histogram by interpreting the density values as a scalar field. This computation is carried out using the Topology Toolkit (\texttt{TTK})~\cite{Tierny2018ttk}. In discrete Morse theory, a critical point of Morse index $i$ corresponds to a critical simplex of dimension $i$~\cite{forman1998morse}. As a result, for 2D scalar fields, a maximum (index 2) is represented by a triangle in the input simplicial complex. Additionally, the $1$-dimensional ascending separatrices connecting saddles to maxima are realized as triangle strips, which increases the geometric complexity of the output. To simplify the representation and ensure that maxima are detected at vertex locations on the original grid, we negate the scalar field prior to computing the MS complex. This inversion transforms the original maxima into minima in the negated field. After the MS complex is constructed, we revert the scalar field to its original form for further analysis. It is important to note that this inversion is a purely technical step to improve numerical stability and does not alter the validity or outcome of the overall method.

\begin{figure*}[!t]
    \centering
   \includegraphics[width=\textwidth]{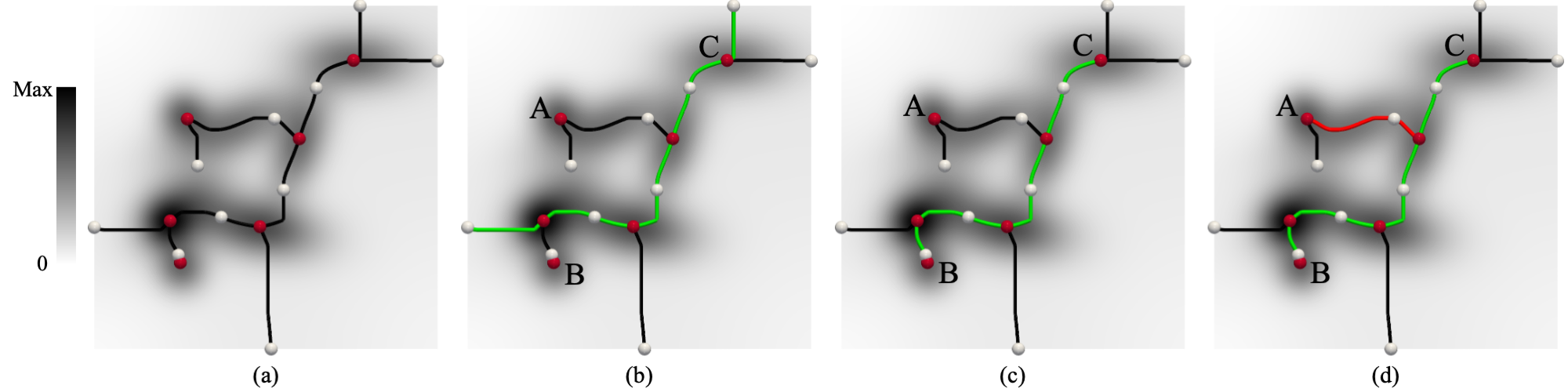}
    \caption{Topology guided path extraction. (a) Minimum spanning tree (MST) of the extremum graph, shown in black. Edge weights are defined as the absolute difference between the scalar values of connected maxima and saddles. (b) The longest path (\ie the diameter) of the MST, highlighted in green. (c) An alternate path, selected interactively to exclude hanging degree-$1$ saddles and include an additional maximum $B$. (d) Two prominent ridge paths are interactively selected, shown in red and green.}
    \label{fig:method_1}
\end{figure*}

\myparagraph{Graph construction.} The maxima of the histogram are assumed to represent prominent features in the data. Our objective is to extract a path that captures the dominant structure in the histogram while preserving these important features. To achieve this, we use the $1$-separatrices of the MS complex, which represent the integral lines connecting saddles to maxima, \ie maximum graph. We then construct a weighted graph in which the nodes consist of all \emph{maxima} and the connected \emph{saddles}, while the edges correspond to the 1-separatrices between them.
%
Let $D \colon \mathbb{R}^2 \to \mathbb{R}$ denote the density field over the histogram\newtext{, is computed via binning as described in \autoref{sec:histselection}}. For an edge connecting a maximum $m$ to a saddle $s$, the edge weight is defined as:
\begin{equation}
w(m, s) = \left| D(m) - D(s) \right|
\end{equation}

 This weighted graph represents a topologically meaningful structure for identifying paths that trace high-density regions and preserve key features in the data. Constructing a minimum spanning tree (MST) on this graph ensures that all important features—represented by the maxima—are retained as nodes, while the most relevant connections between them are preserved. The MST inherently avoids low-density regions, because such regions would correspond to edges with higher weights, \ie greater drops in density. As a result, the extracted path structure aligns closely with the dominant high-density topology of the histogram.
\autoref{fig:method_1}(a) shows the MST computed from the underlying histogram. All prominent maxima are included, and the MST edges (in black) cover the primary high-density structure. Hanging saddles with degree one are also included but can be interactively excluded during fused field computation in later steps, if needed.

\myparagraph{Path extraction.}
After computing the MST, we extract a simple path that captures the major structure of the histogram. Intuitively, the longest path between any pair of nodes, \ie the diameter of the tree, is likely to span a significant portion of the histogram. This path is computed using the same edge weights as the MST, defined by the absolute difference between the scalar values of connected maxima and saddles. However, how well the extracted path captures the key features depends on the placement of critical points and the underlying structure of the histogram. For example, the path may terminate in low-density regions at very low maxima or degree-$1$ saddles, as visible in the green path in \autoref{fig:method_1}(b). In such cases, we interactively filter out vertices located in low-density areas. \autoref{fig:method_1}(c) shows an interactively selected path connecting two maxima, $B$ and $C$. Often, a single path fails to capture multiple meaningful branches, for instance, the branches visiting maxima $A$ and $B$ in \autoref{fig:method_1}(b) are not covered. To address this, individual branches can be manually selected from the multi-branch structure, as shown by the red and green path in \autoref{fig:method_1}(d). The resulting paths provide a summary of important trends or a targeted segment of the histogram while passing through key features.

\subsection{Volume fusion}
\label{sec:volumefusion}
Our next objective is to construct a fused volume with a scalar field $F$, such that a large number of prominent topological features become distinguishable in this volume. The linear path extracted in the previous step serves to separate prominent features when traversed along it. Therefore, we project all features from the histogram onto this path to retain as many important features as possible. It is important to note that some features not intersected by the path may be lost or may overlap with others. However, due to our carefully designed edge weights and the resulting extracted path, we expect to capture most of the significant features.

The extracted path often contains a large number of sharp corners due to the discrete nature of the extracted $1$-separatrices, which can introduce artifacts during projection and lead to noise in the fused volume. To resolve this, we smooth the path by fitting a cubic B-spline using the \texttt{scipy} library. The smoothing factor is chosen carefully to preserve important features. \newtext{Excessive smoothing can flatten key regions, merge features, and reduce projection effectiveness. The chosen smoothing factor for each dataset is reported in \autoref{tab:computeTimes}.} The smoothed spline is sampled densely with $10^6$ equally spaced samples along arc length to minimize discretization errors in the histogram-to-spline mapping. 

\begin{figure*}[!t]
    \centering
   \includegraphics[width=\textwidth]{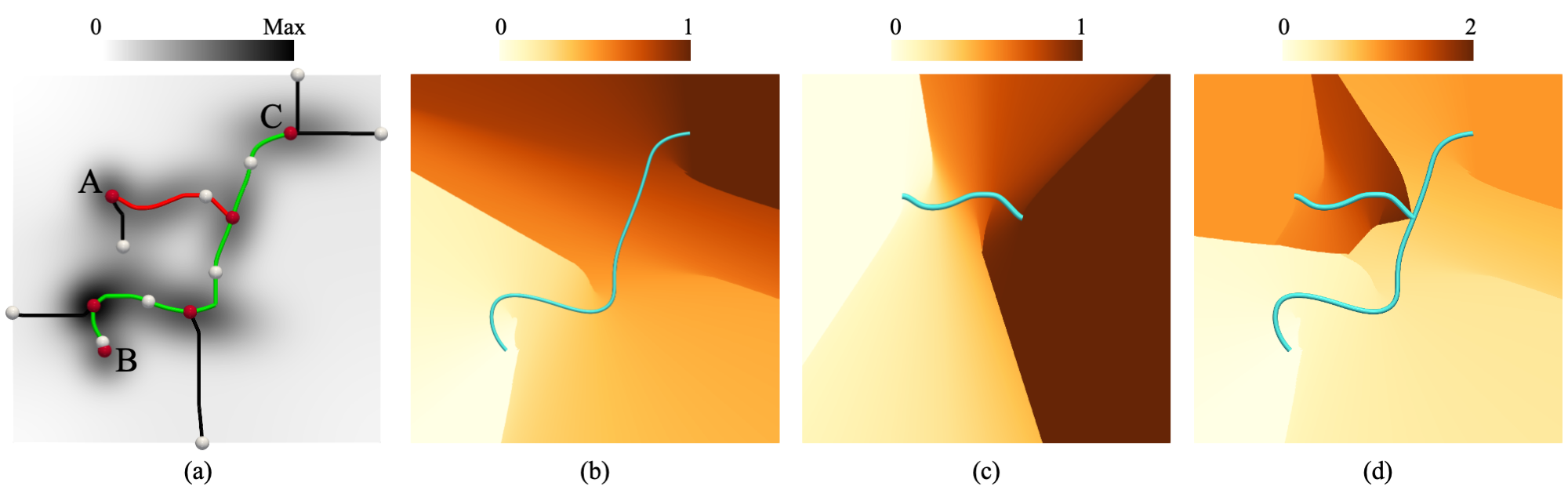}
    \caption{Multi-branch fusion in histogram space. (a) Two prominent ridge paths are interactively selected, shown in red and green. (b) Fitted spline and fused scalar field corresponding to the green path, with values ranging from $0$ to $1$. (c) Fused field generated for the red path, also ranging from $0$ to $1$. (d) Merged scalar field created by combining both paths based on the proximity of histogram points to the respective splines, with values ranging from $0$ to $2$.}
    \label{fig:method2}
\end{figure*}

\begin{figure}[!t]
    \centering
   \includegraphics[width=\columnwidth]{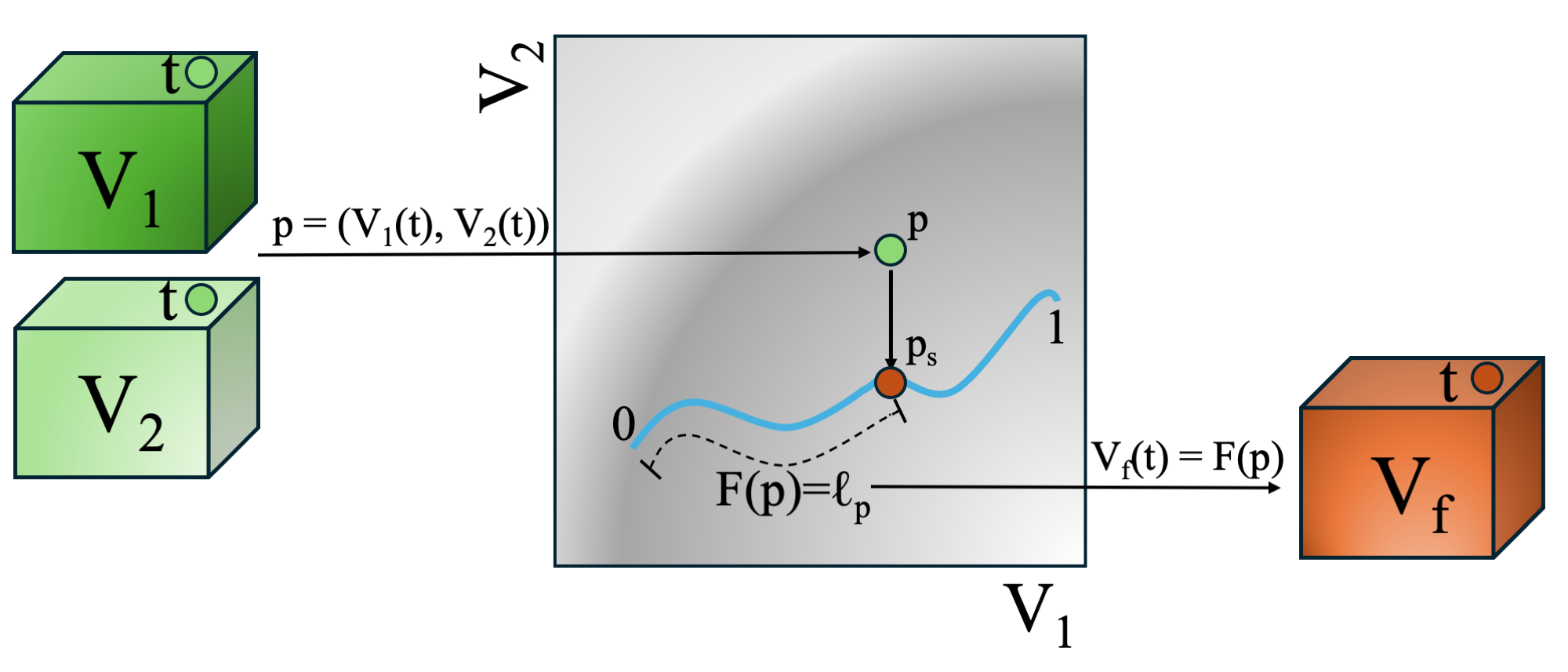}
    \caption{Grid parameterization and fused volume computation. A spatial point $t$ maps to $p = (V_1(t), V_2(t))$ in histogram space, which is projected onto point $p_s$ on the topologically guided spline. The arc length $l_p$ from the spline’s start to $p_s$ parametrizes the histogram and is transferred back to the spatial domain to form the fused volume $V_f$.
} 
    \label{fig:volumeFusion}
\end{figure}

\myparagraph{Grid parameterization.} For each grid point $p$ in the histogram space, we identify the closest point $p_s$ on the spline. We then compute the normalized arc length $\ell_p \in [0, 1]$ from the start of the spline to $p_s$, and assign it as the scalar value to construct the fused volume.
\begin{equation}
F(p) = \ell_p
\end{equation}

\autoref{fig:method2}(a) shows a path with two branches, highlighted in red and green. \autoref{fig:method2}(b) and (c) show the fitted splines and corresponding grid parameterizations for the red and green branches, respectively.

\myparagraph{Multiple branches.} In the case of multiple branches $ \{B_i\}_{i=1}^{n}$ in the path, we construct a separate scalar field $F_i$ for each branch. These branch-wise fused volumes can be analyzed independently, or a single merged histogram-to-spline mapping can be computed as follows: Map a grid point $p$ in the histogram space to all branches and find the closest mapping, say $p_k$, on branch $B_k$. Arc length $\ell_k \in [0,1]$ is computed from start of $B_k$ to $p_k$, and the final scalar value is assigned as:
\begin{equation}
F(p) = k + \ell_k
\end{equation}

\autoref{fig:method2}(d) shows a single merged field ranging from $0$ to $2$, for the two-branch path.
This approach allows us to encode segmentation information from multiple branches into a single scalar field, while preserving topological context that can be used for volume rendering and designing transfer functions.

\myparagraph{Fused volume computation.} Finally, the scalar field computed using the histogram-to-spline projection is transferred back to the spatial domain as illustrated in \autoref{fig:volumeFusion}. Each spatial point $t$ has corresponding scalar values $V_1(t)$ and $V_2(t)$ in the two input volumes, representing a point $p = (V_1(t), V_2(t))$ in the histogram space. The fused scalar field $F$, computed via grid parameterization over the histogram, assigns a scalar value to each such point $p$. The fused volume $V_f$ is then defined as:
\begin{equation}
\forall t:\quad V_f(t) = F(p), \quad \text{where } p = (V_1(t), V_2(t))
\end{equation}
This fused volume is expected to preserve the most prominent joint features of the input volumes that lie along the topologically guided path in the histogram space and is used for subsequent visualization and analysis.
Note that the scalar values of the fused volume lie in the range $[0,1]$, or $[0,n]$ for multiple branches.

\subsection{Fused volume rendering}
Visualizing the fused volume is straightforward.
Standard volume rendering methods can be used for this scalar volume in combination with 1D transfer functions or 2D transfer functions based on the gradient magnitude.
Alternatively, the grid parametrization can be interpreted as a 2D transfer function and sampled accordingly using the voxel values $V_1(t),\ V_2(t)$ of the input volumes during volume raycasting. 
This can be beneficial with respect to memory consumption, depending on the memory layout and data format of the underlying data. \newtext{Although a 2D histogram can directly enable a 2D transfer function, feature identification within it is often challenging. In contrast, the fused scalar field fits naturally into standard 1D workflows, avoids sampling of both high-resolution volumes, and reduces the dimensionality of the data, making 1D transfer functions easier to construct, providing a consistent pipeline for both single and multivolume cases. As a minor benefit, it also reduces memory usage compared to storing high-resolution 2D histograms while consolidating the data into a single volume.}

\begin{figure*}[!t]
    \centering
   \includegraphics[width=\textwidth]{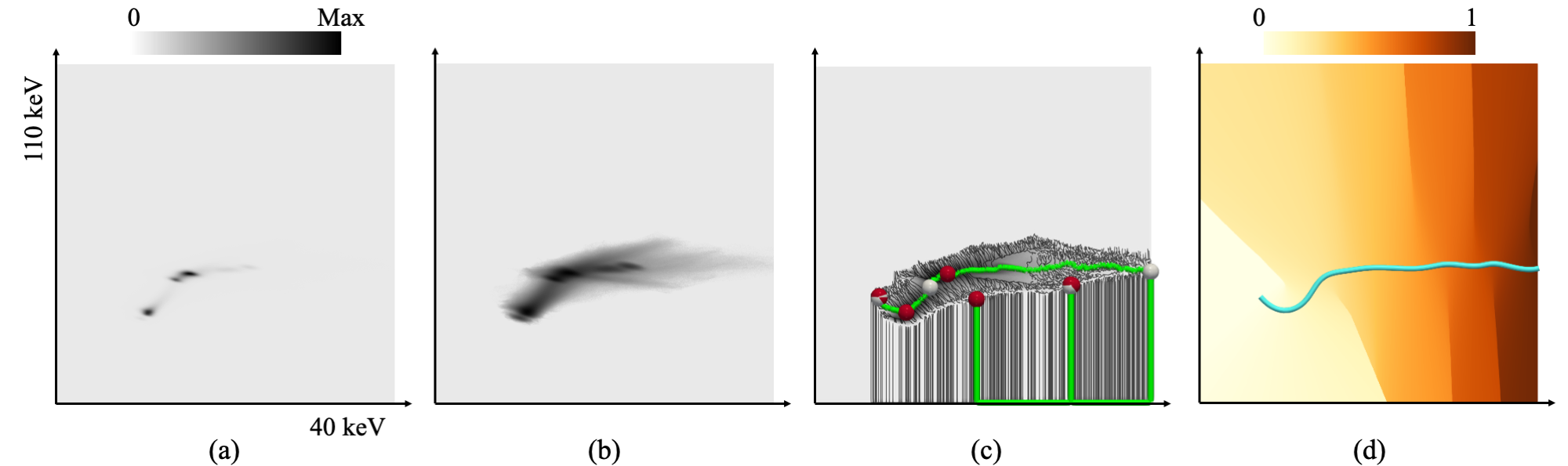}
    \caption{Volume fusion for human heart dataset. (a) Linear scale 2D histogram showing overall density distribution. (b) Log scale 2D histogram highlighting finer features in low-density regions. (c) Noisy MST of the extremum graph overlaid on the simplified scalar field; the longest path is shown with maxima (red) and saddles (white). Endpoints in low-density areas are suppressed interactively. (d) The smoothed spline fit of the longest path is used to project the histogram and generate the fused volume.} 
    \label{fig:heartpipeline}
\end{figure*}

\begin{figure}[!t]
    \centering
    \subfloat[40\,keV]{\includegraphics[width=0.3\columnwidth]{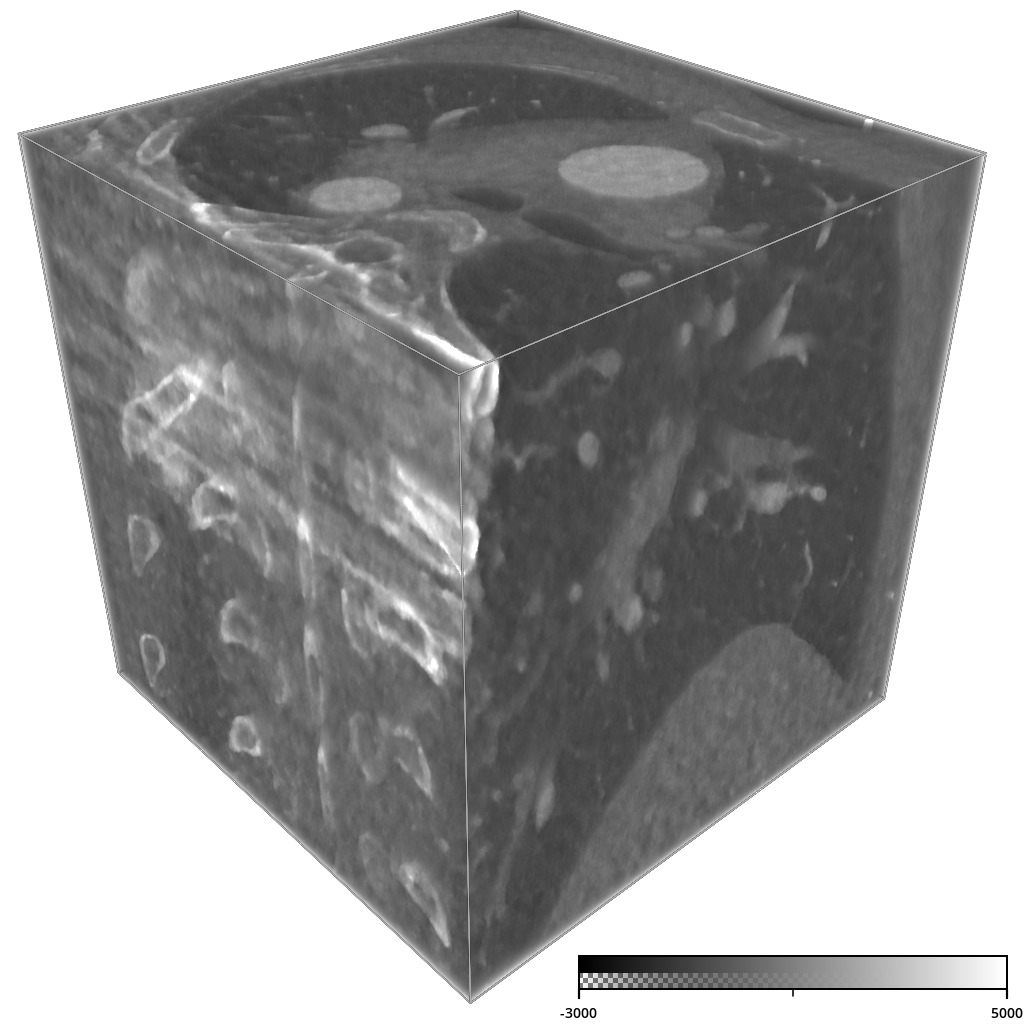}}%
    \hfill\subfloat[110\,keV]{\includegraphics[width=0.3\columnwidth]{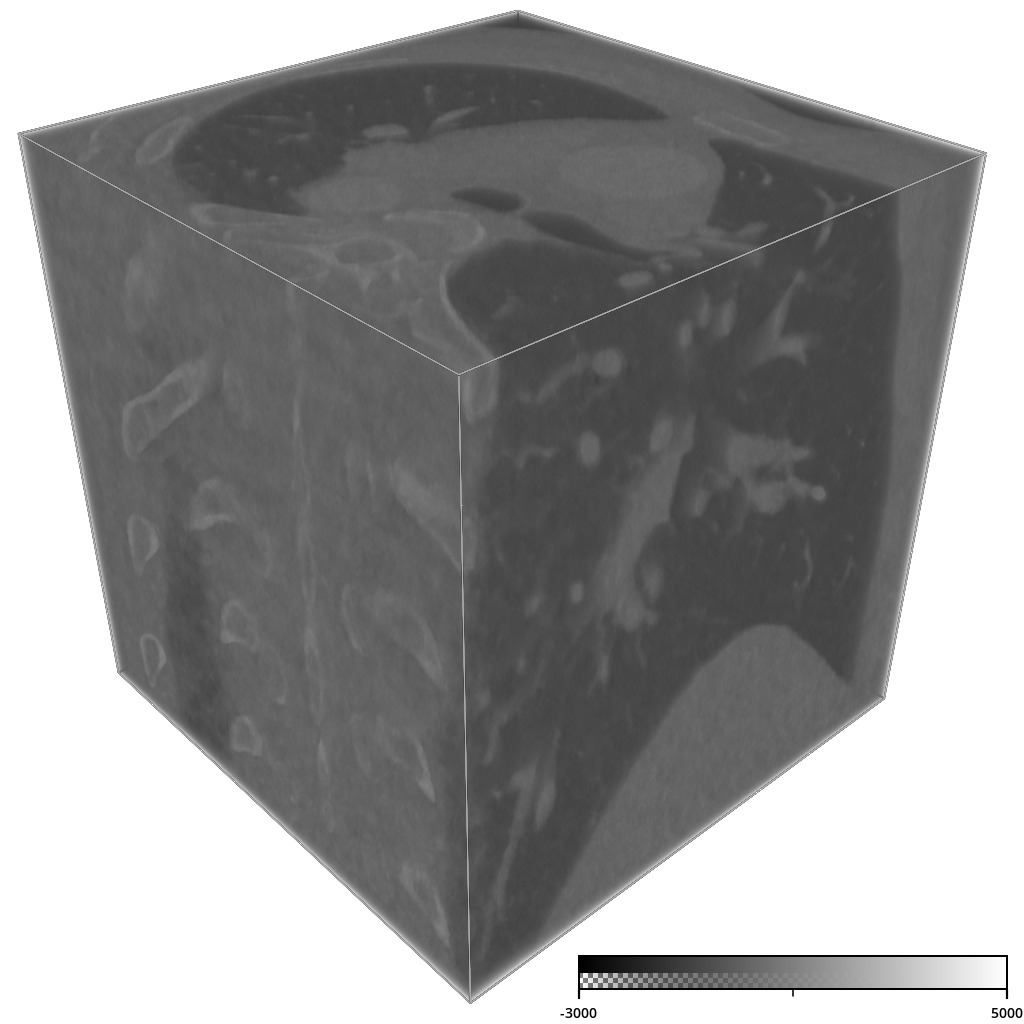}}%
    \hfill\subfloat[fused volume]{\includegraphics[width=0.3\columnwidth]{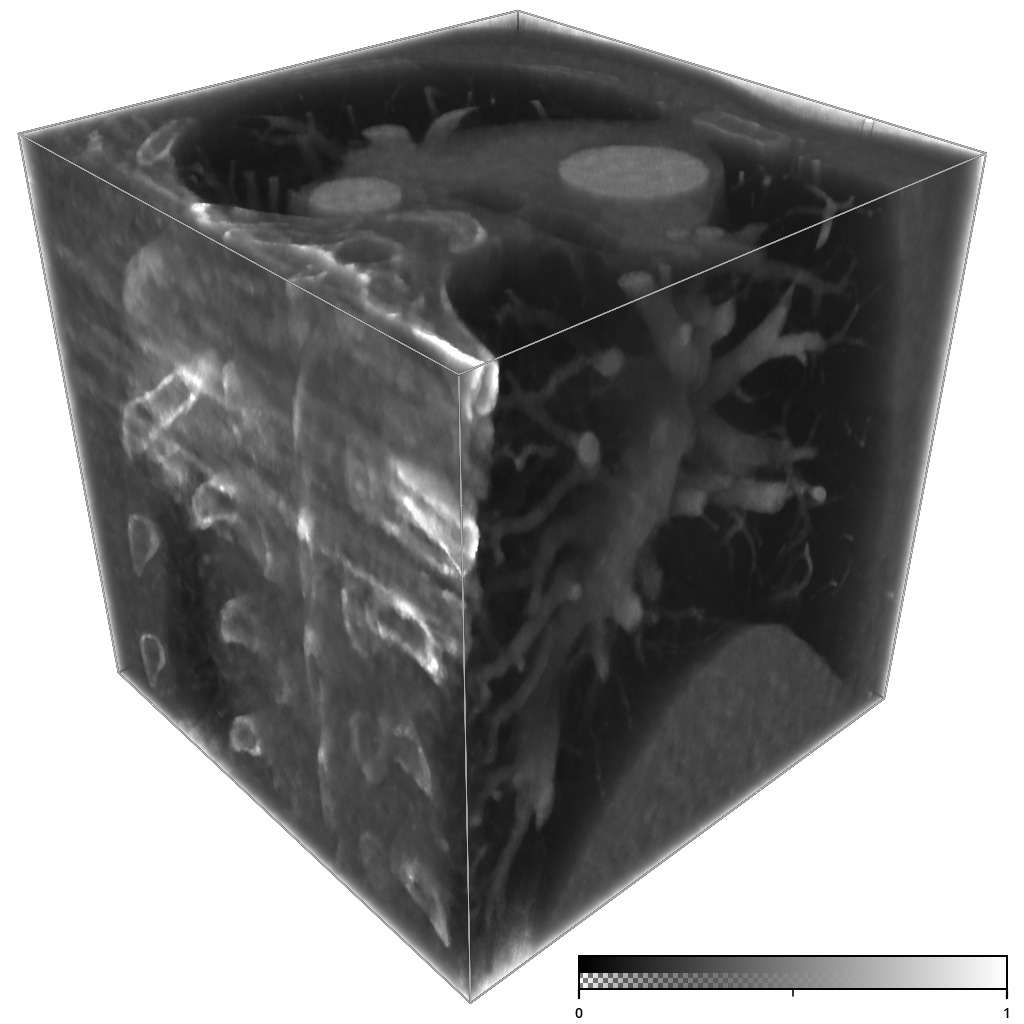}}%
    \\
    \subfloat[]{\includegraphics[height=2.4cm]{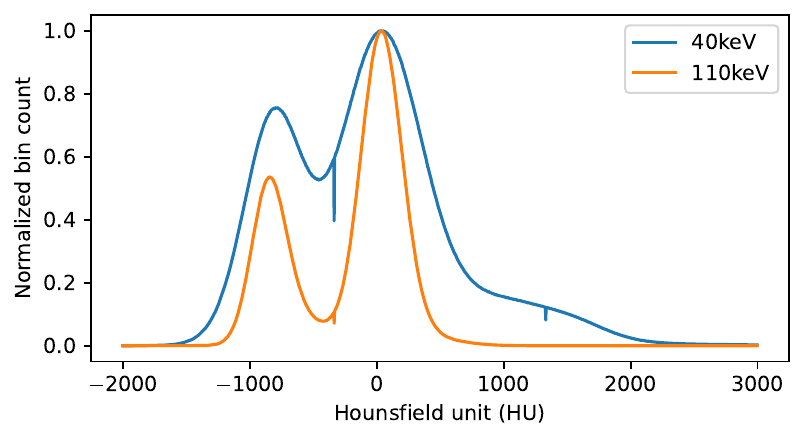}}%
    \hfill\subfloat[]{\includegraphics[height=2.4cm]{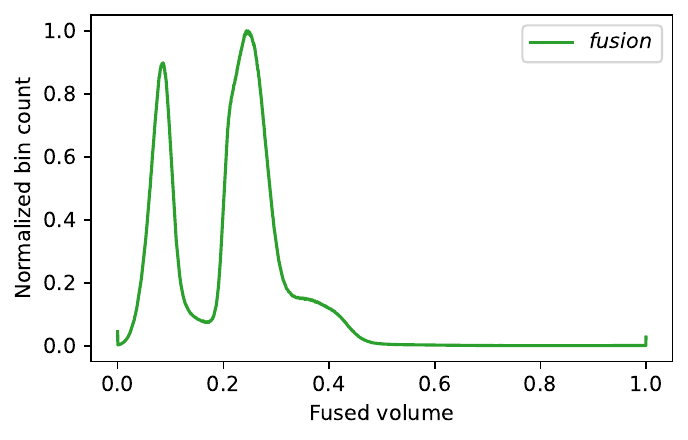}}
    \caption{Volume rendering of the human heart dataset.
         (a)-(c) A basic grayscale transfer function is applied to the input volumes for 40\,keV and 110\,keV, and the fused volume.
        (d)-(e) The 1D histograms corresponding to the underlying scalar fields.
    } 
    \label{fig:heartoutcome}
\end{figure}

\section{Results}
\label{sec:results}
In this section, we demonstrate the effectiveness of our method on one synthetic dataset and two real-world medical datasets. Both the real datasets were acquired using a dual-source PCCT scanner (NAEOTOM Alpha, Siemens Healthcare) in ECG-triggered cardiac dual-source spiral mode, enabling spectral post-processing. The scan parameters included a rotation time of $0.25$\,s, temporal resolution of $66$\,ms, collimation of $144 \times 0.4$\,mm, and a pitch of $0.31$. The phantom scan (\autoref{results_lamb_heart}) used a synthetic ECG signal set to $60$\,bpm. Spectral images were reconstructed at varying virtual monoenergetic image (VMI) energy levels.

All experiments are conducted on a machine with a $5.0$ GHz AMD Ryzen Threadripper $7960$X processor ($24$ cores) and $64$\,GB of RAM. Visualizations are generated using ParaView~\cite{Ayachit2015ParaView} and Inviwo~\cite{inviwo2019}\newtext{; the latter is specifically used to produce the volume renderings shown in \autoref{fig:heartoutcome}, \autoref{fig:lambheartpipeline}, and \autoref{fig:lambresults}.} The extremum graph is computed using the Topology ToolKit \texttt{(TTK)}~\cite{Tierny2018ttk}. Dataset-specific parameters and computation times are summarized in \autoref{tab:computeTimes}. The computation time scales proportionally with the histogram resolution, as the most computationally intensive step involves mapping every point in the histogram space onto the spline during grid parametrization.

\subsection{Circular Gaussians}
A synthetic dataset having multiple Gaussians arranged along a circular path is shown in \autoref{fig:teaser}(a). This dataset represents the attribute space—or 2D histogram—of two scalar fields. These scalar fields may either correspond to two distinct attributes defined over the same spatial domain or to the same attribute measured under different conditions, such as in multi-energy PCCT datasets. Each Gaussian corresponds to a distinct feature and is easily identifiable in the 2D histogram. However, extracting these features using either of the two fields independently is challenging: due to their overlapping nature, multiple Gaussians cannot be cleanly separated using isosurfaces from just one field. This is visible in the 1D histograms obtained by projecting the density from the 2D attribute space onto the individual axes—only $3$ prominent peaks are visible, despite the presence of $8$ underlying bivariate features. Moreover, dimensionality reduction techniques such as principal component analysis (PCA) are also ineffective in this case. The 2D histogram has rotational symmetry and, therefore, lacks any dominant principal direction, making it difficult for PCA to align with and separate the feature peaks along any principal axis. \autoref{fig:teaser}(b) shows the Morse–Smale complex computed in the attribute space. \autoref{fig:teaser}(c), the green path represents the MST restricted to maxima and saddles, which also corresponds to the longest path (or diameter) of the tree. This path effectively traverses all major features in the data. A smooth spline is then fitted along this path, as illustrated in \autoref{fig:teaser}(d). Each point in the histogram is projected onto its nearest point on the spline, assigning it a parameter value between $0$ and $1$, based on its relative arc length from the starting point. This parameterization enables the construction of a fused scalar field in the original spatial domain. To evaluate whether this fused field captures all significant features, we compute a histogram by aggregating the density values of all grid points projected onto each location along the spline. This yields a one-dimensional density distribution over the spline, shown in \autoref{fig:teaser}(d). The histogram reveals $8$ distinct peaks, corresponding to identifiable features.

\subsection{Human heart}
\label{sec:humanheart}
This PCCT dataset was drawn from a larger collection of ECG-gated cardiac CT angiography scans performed during routine clinical care and approved for retrospective research use. The dataset consists of volumes captured at different energy levels ranging from $40$\,keV to $110$\,keV in increments of 10\,keV. The goal is to extract distinct segments corresponding to key anatomical and material features such as the heart, lungs, air, soft tissue, and contrast agent in the blood. A single-energy volume often fails to clearly identify the boundaries of these features, which is why multiple energy levels need to be reconstructed.
After examining histograms for all pairs of energy levels, we selected the histogram corresponding to the $40$\,keV and $110$\,keV volumes due to its well-defined structure and visibly prominent features and the lowest correlation of $0.692$. 
Directly neighboring energy levels in contrast have a high correlation ranging from $0.98$ to $0.997$.
This high correlation is reflected by the almost perfectly diagonal lines in the 2D histograms (cf.\ histogram matrix in \autoref{fig:pipeline}). 

\autoref{fig:heartpipeline}(a) shows the selected 2D histogram. Only a few key maxima are identifiable, while many features appear faint due to their low values. Mapping the histogram to a logarithmic scale reveals the rest of the structure and more prominent features, as shown in \autoref{fig:heartpipeline}(b). 
To reduce noise, the histogram is simplified using topological persistence. The parameter choices for simplification are provided in \autoref{tab:computeTimes}. The resulting MST, overlaid in black on the simplified histogram, is shown in \autoref{fig:heartpipeline}(c). While the MST is still noisy--—particularly towards the right side of the histogram---the longest path in the tree, highlighted in green, effectively captures the primary structure. This path passes through key maxima associated with important regions but also includes a few extra points in a low-density region. To extract the meaningful portion of the path, we discard these low-density points and retain only the segment that lies within a high-density region. \autoref{fig:heartpipeline}(d) illustrates the smoothed spline fitted to this refined path, along with the parameterized grid obtained by projecting grid points onto the spline. This parameterized representation is then mapped back to the spatial domain to generate a fused volume for further analysis.

\autoref{fig:heartoutcome} depicts the two input volumes and the resulting fused volume, along with corresponding 1D histograms where the fused volume retains the combined features of the two input volumes.
The decreasing slope on the right side of the second peak between 800\,HU and 2000\,HU at 40\,keV becomes more emphasized in the fused volume.
There is also a noticeable difference in the leading slope right before the second peak, which could indicate a difference in material or an artifact due to spline fitting and grid parametrization. 
Overall, the fused volume increases feature contrast compared to the two individual input volumes while maintaining their features at the same time.

\begin{figure*}[!t]
    \centering
   \includegraphics[width=\textwidth]{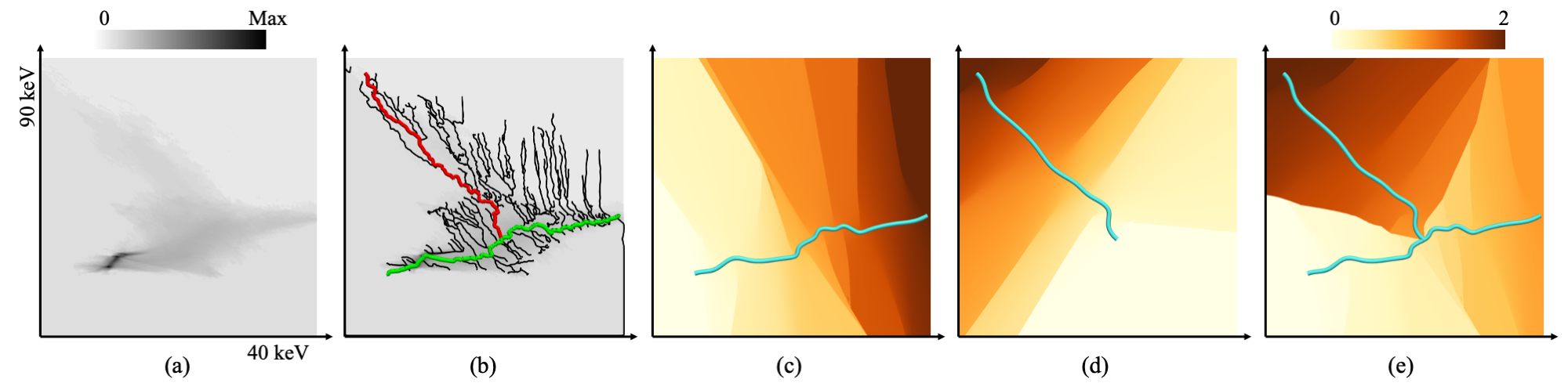}\\
   \addtocounter{subfigure}{5}
   \subfloat[40\,keV]{\includegraphics[width=0.19\textwidth]{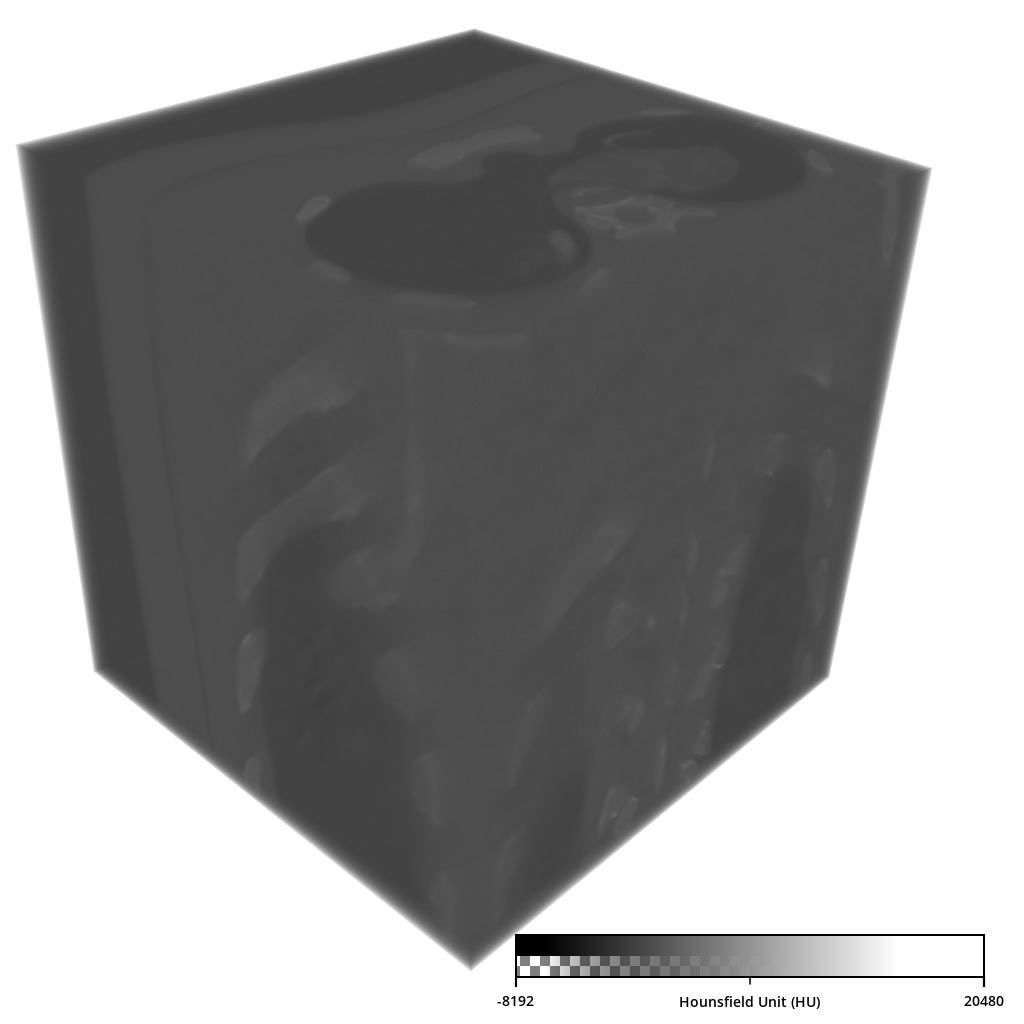}}%
   \hfill\subfloat[90\,keV]{\includegraphics[width=0.19\textwidth]{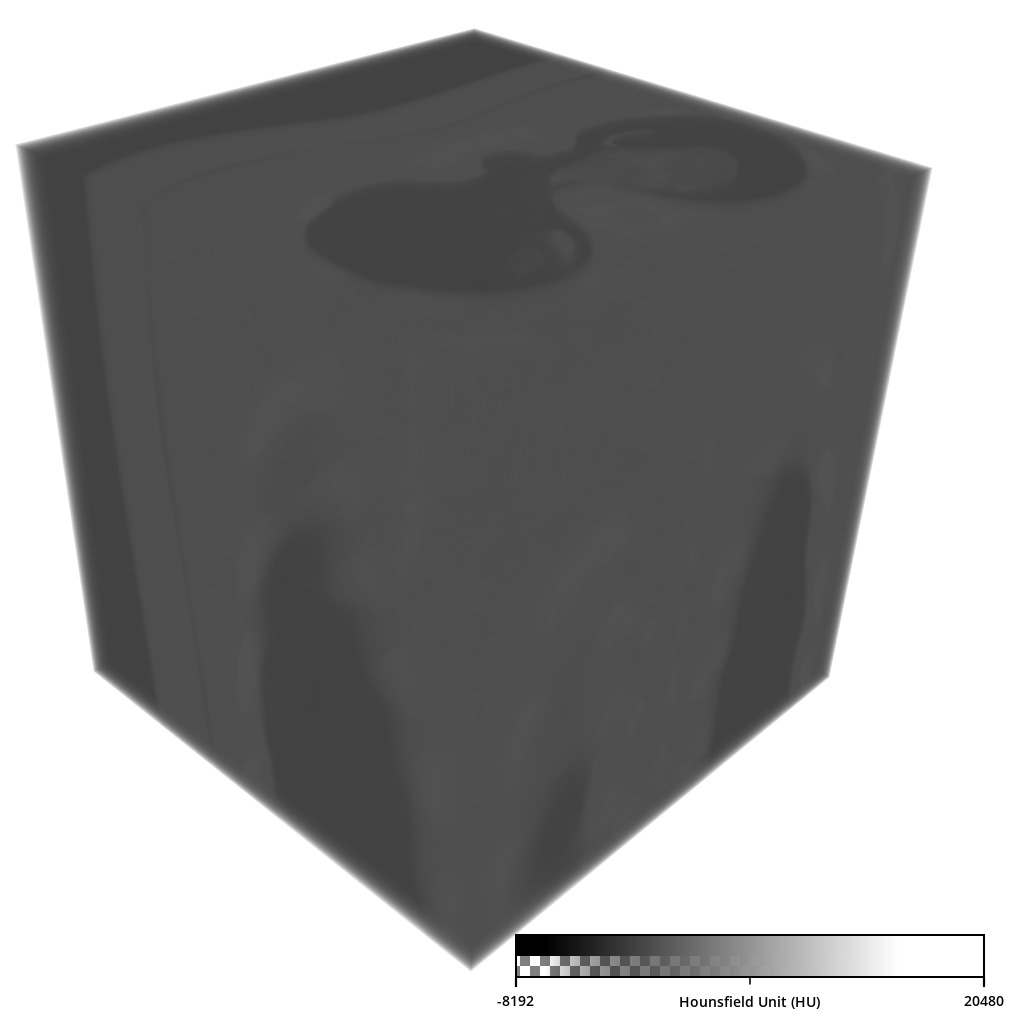}}%
   \hfill\subfloat[fused volume (c)]{\includegraphics[width=0.19\textwidth]{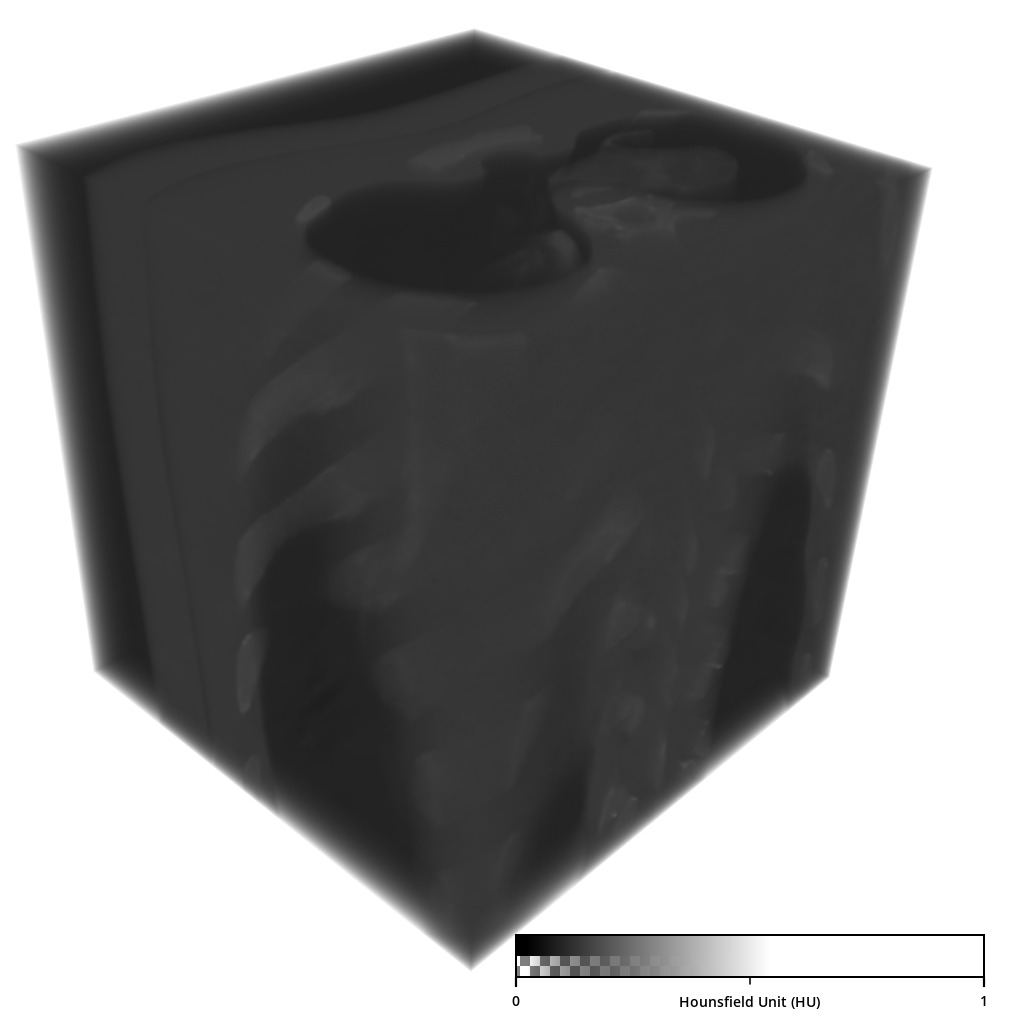}}%
   \hfill\subfloat[fused volume (e)]{\includegraphics[width=0.19\textwidth]{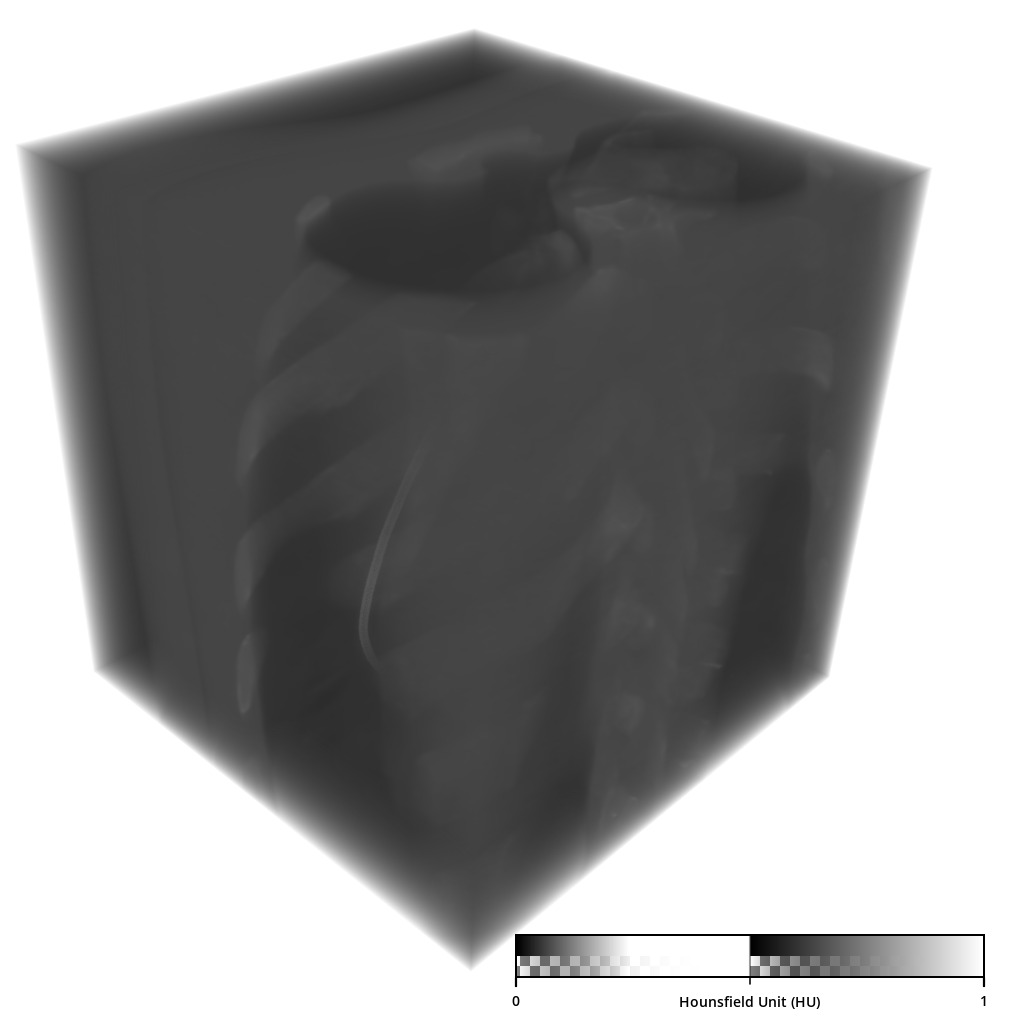}}%
   \hfill\subfloat[]{\includegraphics[width=0.19\textwidth]{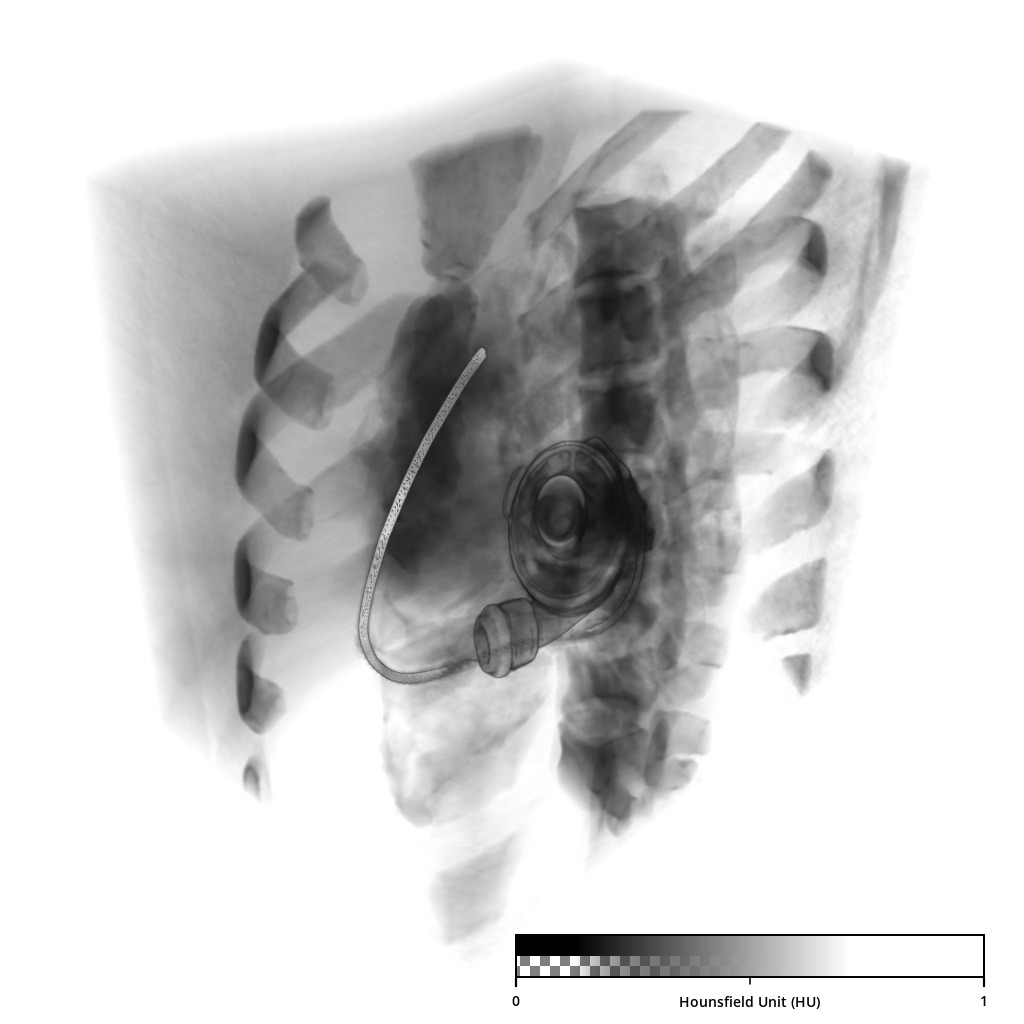}}%
    \caption{Phantom lamb heart dataset.
    Volume fusion (a)-(e) and volume rendering (f)-(j). (a) Log scale histogram exhibiting undesired scatter caused by the presence of an artificial metallic heart. (b) Two prominent paths, shown in red and green, are interactively selected to extract branching structures within the histogram. Three fused volumes are generated: (c) grid parameterization along the green path (used in (h)), (d) along the red path, and (e) the combined parameterization merging both paths (used in (i)).
    (j) slightly adjusted transfer function for fused volume (c).
    } 
    \label{fig:lambheartpipeline}
\end{figure*}

\begin{figure*}[!t]
    \centering
    \includegraphics[width=0.22\textwidth]{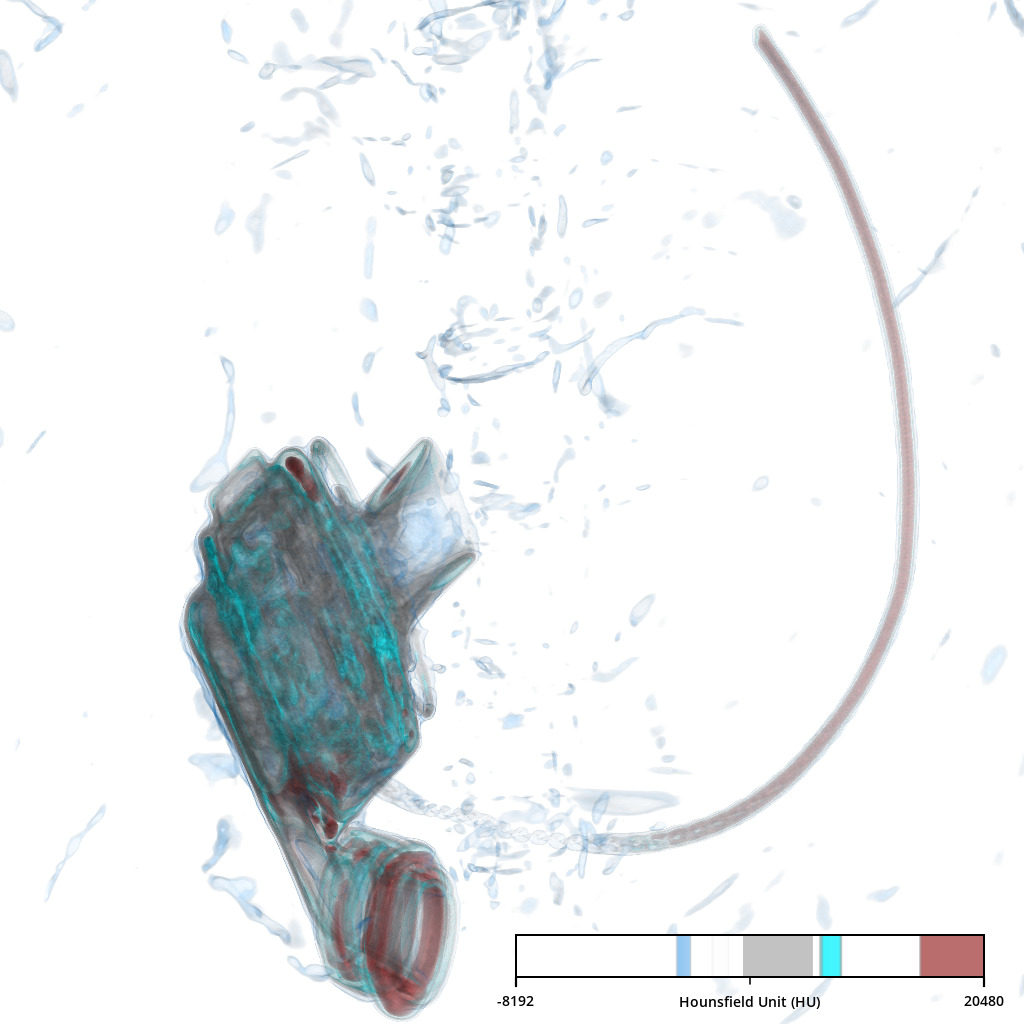}%
    \hfill\includegraphics[width=0.22\textwidth]{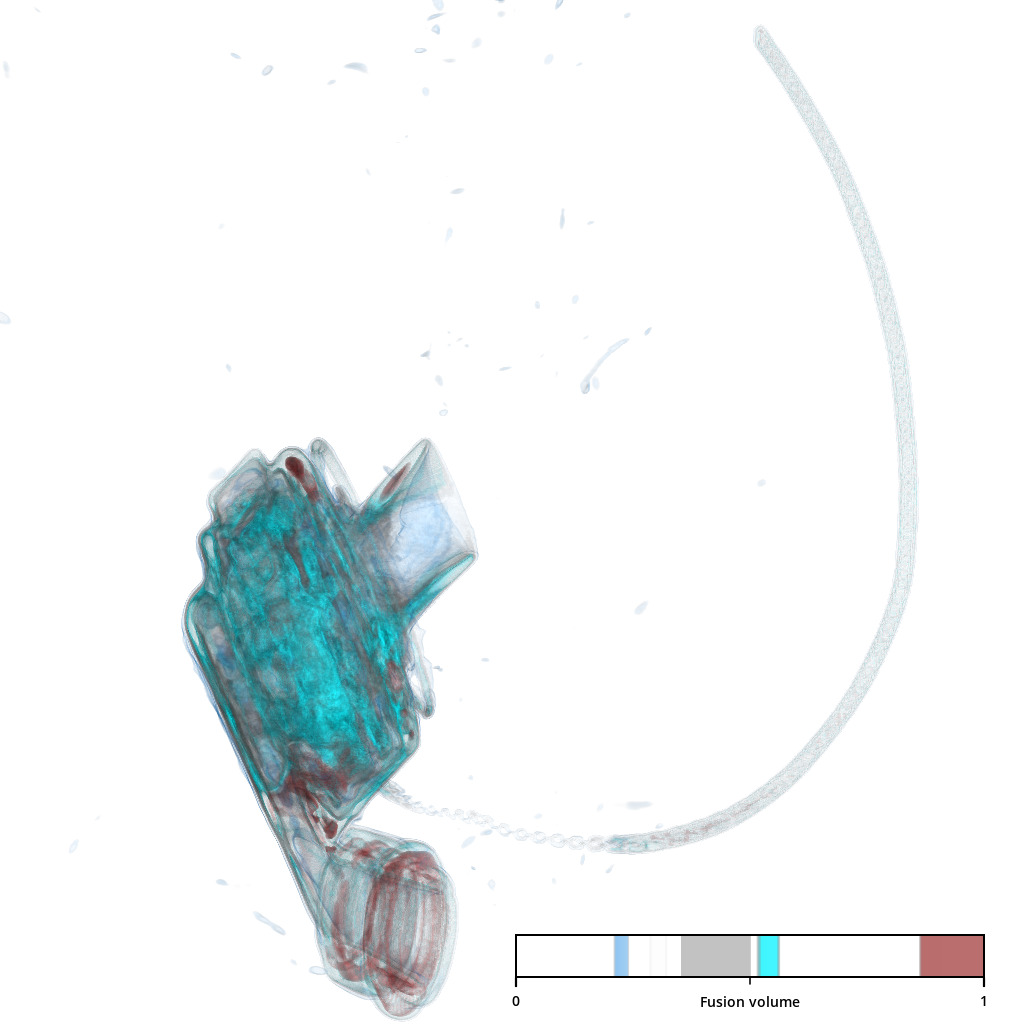}%
    \hfill\includegraphics[width=0.22\textwidth]{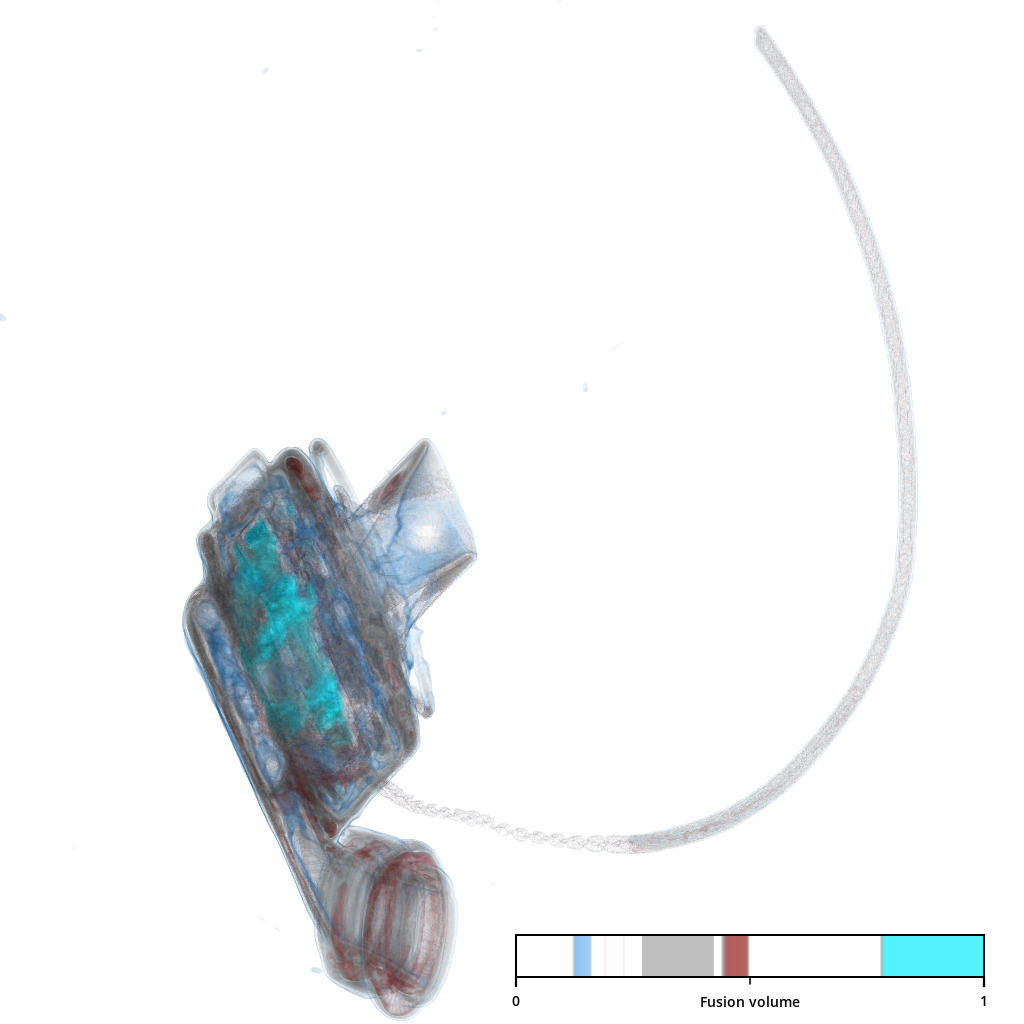}%
    \hfill\includegraphics[width=0.22\textwidth]{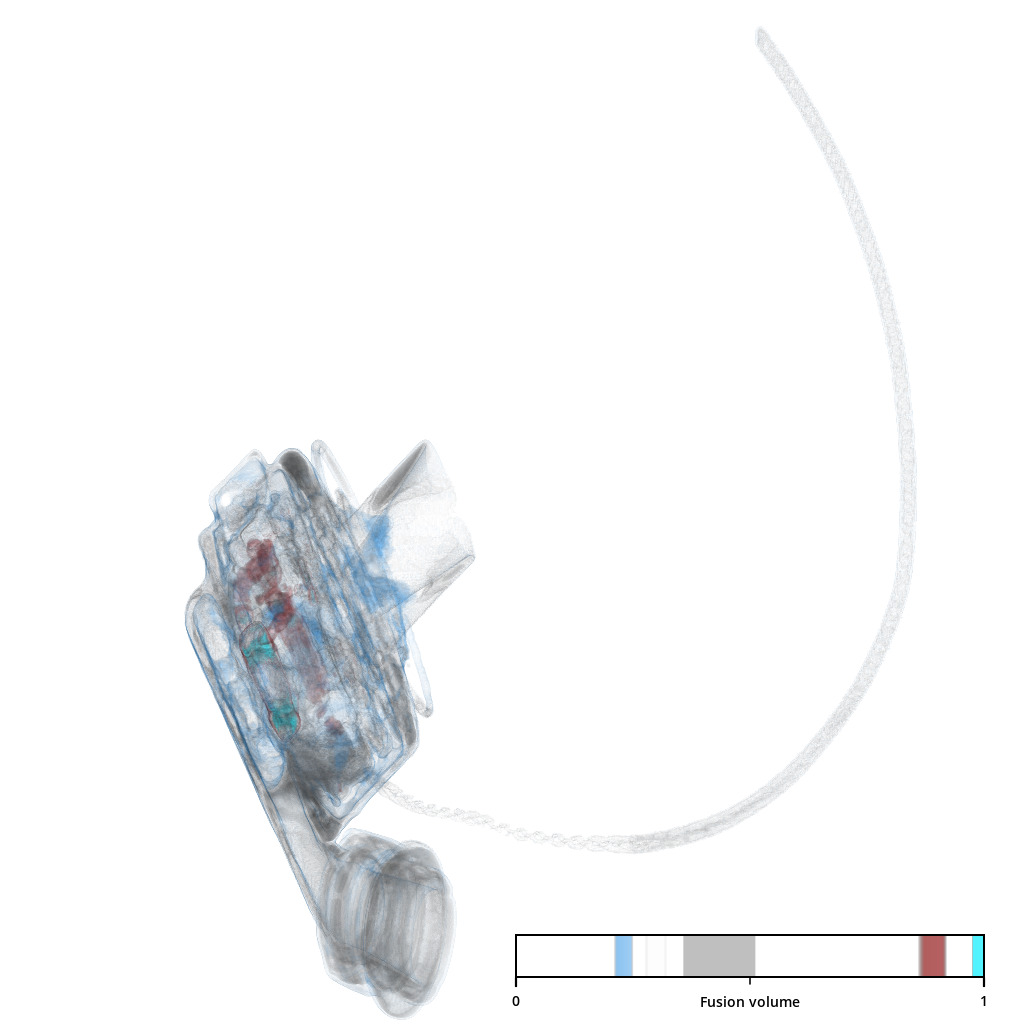}\\
    \subfloat[40\,keV]{\includegraphics[width=0.22\textwidth]{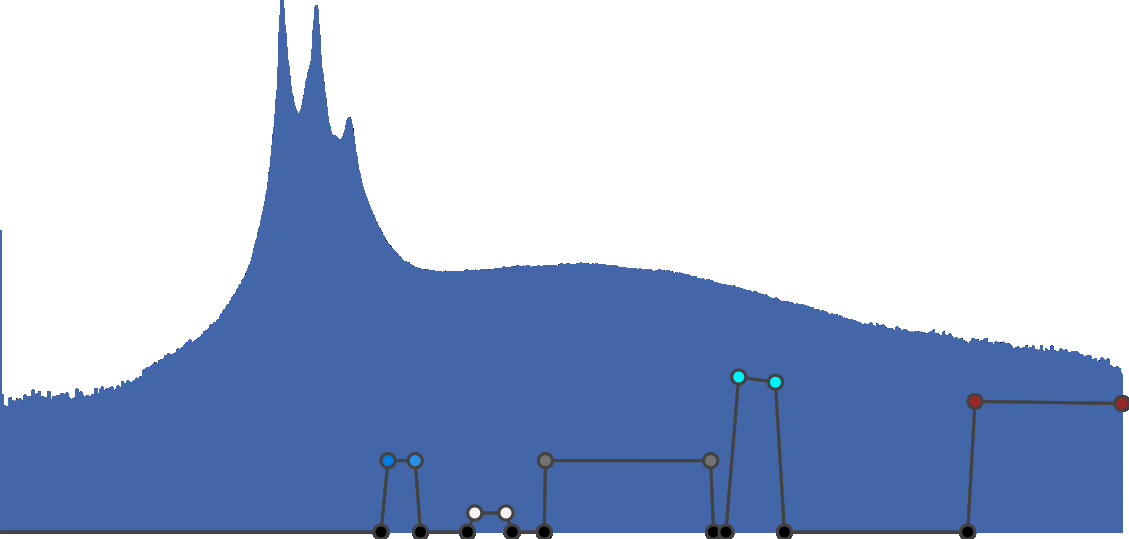}}%
    \hfill\subfloat[fused volume (green path)]{\includegraphics[width=0.22\textwidth]{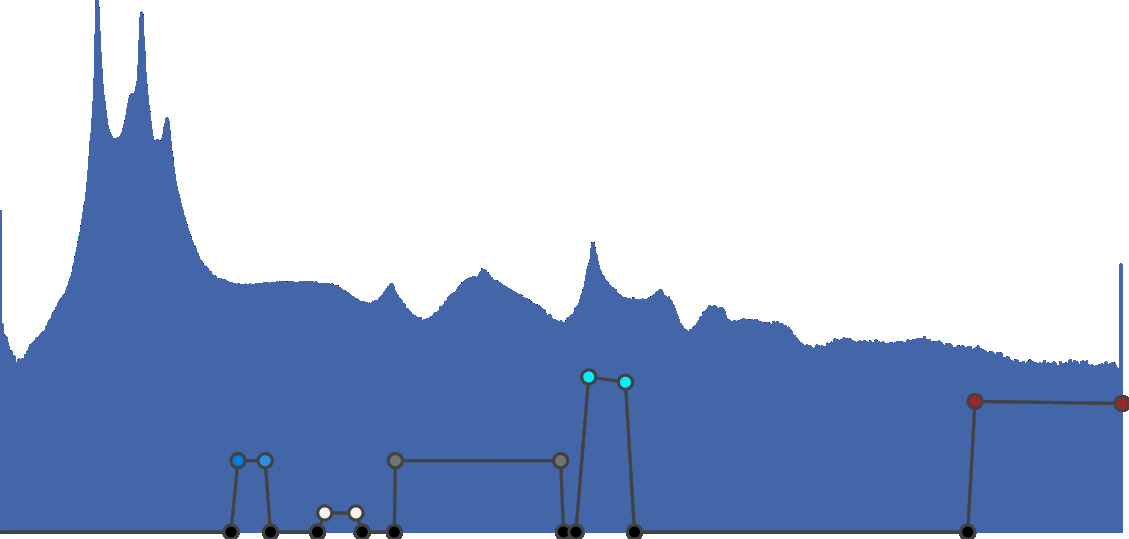}}%
    \hfill\subfloat[fused volume (both paths)]{\includegraphics[width=0.22\textwidth]{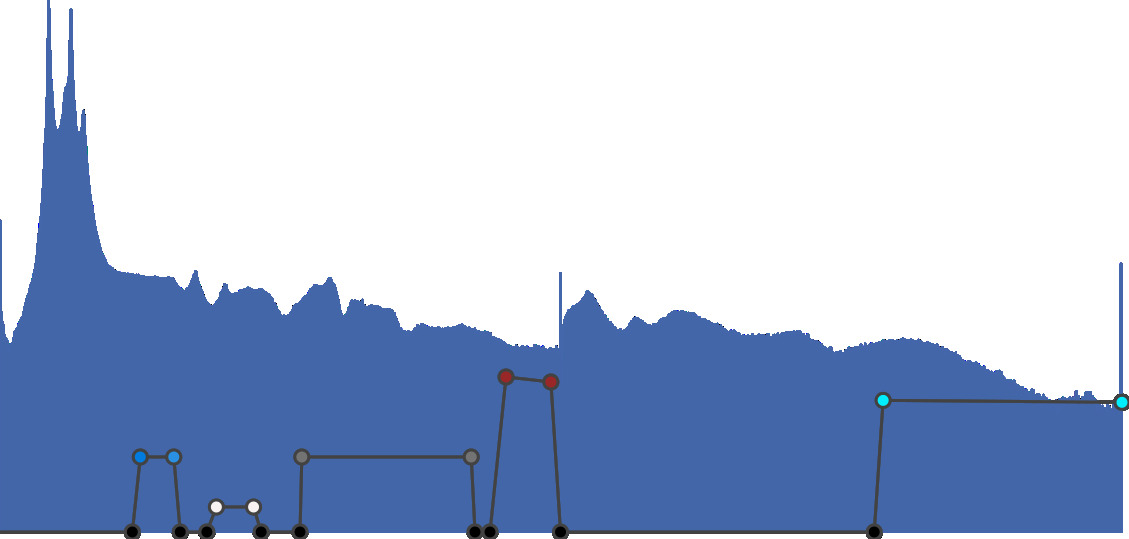}}%
    \hfill\subfloat[fused volume (both paths)]{\includegraphics[width=0.22\textwidth]{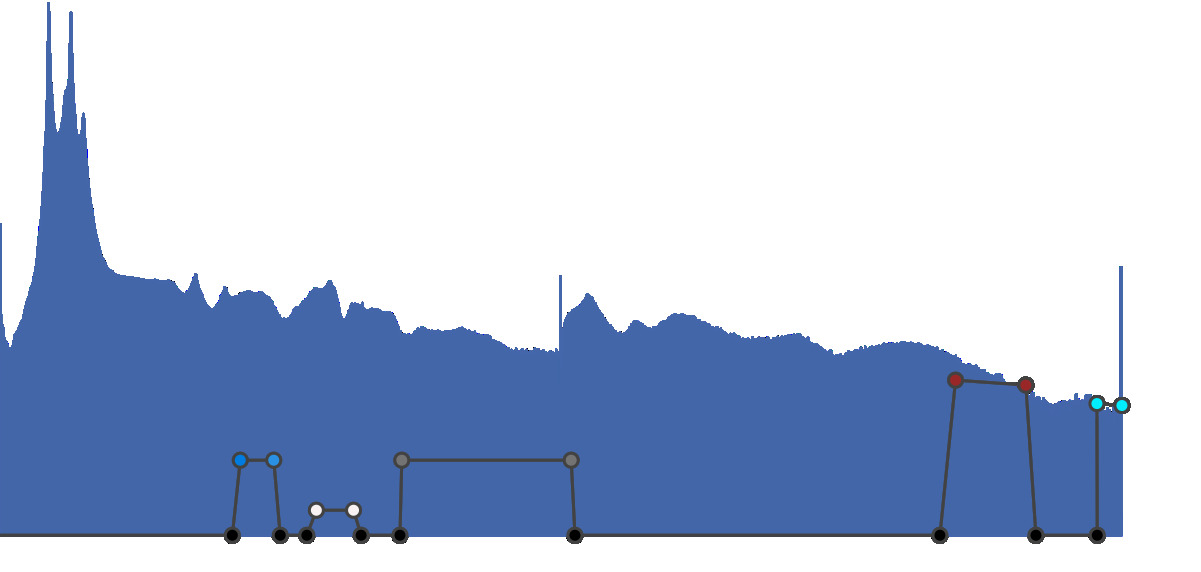}}\\
    \caption{Volume rendering of the artificial heart in the phantom lamb heart dataset.
    (a)-(b) The fused volume results in less clutter with an almost identical transfer function.
    (c)-(d) The 1D histogram of the fused volume reveals more features and thereby eases the transfer function design.
    }
    \label{fig:lambresults}
\end{figure*}

\begin{table*}[htbp]
\centering
\caption{Compute times and parameters for different datasets. Times are averaged over five runs.}
\resizebox{\textwidth}{!}{
\begin{tabular}{l c cc cc}
\toprule
\textbf{Dataset} & \textbf{Resolution} &
\multicolumn{2}{c}{\textbf{Parameters}} &
\multicolumn{2}{c}{\textbf{Compute Times (s)}} \\
 & & \textbf{Persistence Threshold} & \textbf{Smoothing Factor} & \textbf{Simplification \& MS Complex} & \textbf{Graph Computation to Volume Fusion} \\
\midrule
Circular Gaussians         & 1000×1000 & 0.00 & 0.01 & 0.3146 & 21.8730 \\
Human Heart                & 2048×2048 & 0.15 & 0.10 & 1.2058 & 54.8434 \\
Phantom Lamb Heart (Green) & 989×989   & 0.10 & 0.05 & 0.7386 & 26.8626 \\
Phantom Lamb Heart (Red)   & 989×989   & 0.10 & 0.05 & 0.7466 & 27.0240 \\
\bottomrule
\end{tabular}
}
\label{tab:computeTimes}
\end{table*}

\subsection{Phantom lamb heart}
\label{results_lamb_heart}
This multi-energy PCCT scan of a phantom lamb heart was originally scanned to study metal artifacts. In this setup, a lamb’s heart and lung, into which a HeartMate $3$ LVAD prototype was sutured at the left ventricular apex, were inserted into a commercial chest phantom (Multipurpose Chest Phantom N1, Lungman, PH-1; Kyoto-Kagaku Co., Ltd., Japan) with added chest plates. The ventricles were filled with iodine contrast agent, and simulated tumors ($3.5$\,mm and $8$\,mm) with distinct Hounsfield units were placed in the cavity. Further details about the phantom can be found in work by Konst~\etal~\cite{Konst2024PCCT}.

For this dataset, we select the 2D joint histogram constructed from the $40$\,keV and $90$\,keV volumes. After the topological simplification via persistence (parameter settings are listed in \autoref{tab:computeTimes}), the resulting histogram is shown in \autoref{fig:lambheartpipeline}(a). This dataset is particularly challenging because of the metallic heart, which produces pronounced artifacts and intensity scatter along the vertical axis, clearly visible in the histogram. 
The scatter disturbs the main structure and complicates the extraction of a clean path through the key features. \autoref{fig:lambheartpipeline}(b) overlays the MST of the extremum graph in black on the simplified histogram. Two dominant ridges, highlighted in red and green, capture the features in the prominent directions. We interactively select each ridge and compute separate parameterized grids (\autoref{fig:lambheartpipeline}(c) and (d)), yielding two fused volumes. To derive a single fused volume that preserves the global structure while differentiating the regions associated with the red and green ridges, we construct a combined parameterization as described in Section~\ref{sec:volumefusion} (\autoref{fig:lambheartpipeline}(e)).

When focusing on the metallic heart, transfer function design is guided by the features in the 1D histograms of the fused volumes, see \autoref{fig:lambresults}.
In contrast to a single energy level, applying a nearly identical transfer function to the fused volume depicts similar features of the heart while reducing surrounding clutter at the same time.
Considering both paths for the fused volume enables more fine-tuning of the transfer function and different materials of the heart, especially since the red path covers large parts of the scatter region caused by the metal parts (\autoref{fig:lambresults}(c,d)).

\section{Limitations}
\label{sec:limitations}
The proposed method selects the histogram corresponding to the pair of volumes that are least correlated or dependent, uses it for topology-guided path extraction, and volume fusion. The information from the remaining volumes is discarded. The discarded volumes may independently, or in combination, help identify certain material boundaries. \newtext{To address this limitation, an interactive view could display the 2D histogram for any chosen pair of volumes, with the computed path and corresponding features overlaid, allowing users to make an informed selection. Ultimately, our goal is to maximize the use of information from all volumes simultaneously.} This motivates the need for computing an n-dimensional extremum graph and extending each stage of the fusion pipeline to operate in higher dimensions.

Furthermore, the scalar field used for extremum graph computation is the histogram density. The computation of this histogram is crucial. A coarsely approximated or noisy histogram may yield an extremum graph with either no dominant path or an overly jagged one. Although persistence-based simplification can reduce noise to some extent, the resulting structure may still vary with histogram resolution. A finer resolution generally improves accuracy but at the cost of increased computational time, as shown in \autoref{tab:computeTimes}.

The path extraction step requires user interactivity when the automatically computed longest path fails to capture prominent features, as discussed in \autoref{sec:pathExtraction}. \newtext{Additionally, features not intersected by the chosen path may be projected onto segments already containing other features, causing multiple features to share the same scalar values in the fused field. This loss of information could, in principle, be quantified by comparing the number of maxima and their persistence values in the original 2D histogram with those in the fused field; however, accurately computing this loss is a complex task and lies beyond the scope of this work.} Handling multi-branch paths is even more challenging due to user interaction and the introduction of hard boundaries between points mapped to different branches during the grid parameterization step. While the extracted path is expected to represent key structures in the histogram, the grid parameterization step depends purely on geometric proximity between histogram points and the spline. This discards the gradient flow of the density field, leading to a geometric approximation. \newtext{Furthermore, excessive smoothing can cause important features to overlap when projected onto the spline, as mentioned in \autoref{sec:volumefusion}, and requires further study to evaluate its impact.} Although the current results are promising, they could be further improved with topology-aware mapping strategies\newtext{, and a smooth transition of features across branches in multi-branch cases remains an open problem for future exploration.}

Another limitation, especially in the medical context, is that it is no longer possible to directly interpret the scalar values of the fused volume due to the spline projection.
Whereas the underlying individual volumes, in contrast, directly relate to Hounsfield units as used by radiologists.

\section{Conclusion}
\label{sec:conclusions}
This paper introduces a novel topology-guided volume fusion pipeline designed to handle multivolume data acquired through photon-counting computed tomography. By parameterizing the histogram space along a path extracted using topological analysis, the method enables the computation of a fused volume well-suited for transfer function design and volume rendering, minimizing the need to work with multiple input volumes. Results on both synthetic and real datasets demonstrate the effectiveness of the approach. However, incorporating topological information into the histogram-to-spline mapping and utilizing additional volumes could further enhance the quality of the fused volume. \newtext{Automating path extraction through optimization methods that maximize feature coverage or persistence could reduce user interaction, although in certain contexts, interactive control remains valuable for domain experts to target clinically relevant features.} In future work, we plan to address current limitations, particularly by generalizing the pipeline to higher dimensions, integrating topological information into the mapping process, and reducing user interaction to enable automatic fusion for multi-branch paths. 

\acknowledgments{
   This work was supported by the Wallenberg AI, Autonomous Systems and Software Program – Data-Driven Life Science (WASP-DDLS) project “HUDI: Huge Complex Diagnostic Imaging Data: Towards personalized models in the clinical workflow” funded by the Knut and Alice Wallenberg Foundation. The authors acknowledge additional funding support from Swedish e-Science Research Center (SeRC) and the Swedish Research Council (VR) grants 2019-05487 and 2023-0480.
}
\bibliographystyle{abbrv-doi}

\bibliography{main}

\begin{thebibliography}{10}

\bibitem{Alghamdi2023}
R.~Alghamdi, T.~M{\"u}ller, A.~Jaspe-Villanueva, M.~Hadwiger, and F.~Sadlo.
\newblock Doppler volume rendering: A dynamic, piecewise linear spectral representation for visualizing astrophysics simulations.
\newblock {\em Computer Graphics Forum}, 42(3), 2023.

\bibitem{Carolina-Alves2024}
A.~C. Alves, A.~ferreira, G.~Luljten, J.~Kleeslek, B.~Puladi, J.~Egger, and V.~Alves.
\newblock Deep {PCCT}: Photon counting computed tomography deep learning applications review.
\newblock {\em arXiv:2402.04301v1 [eess.IV]}, 2024.

\bibitem{Ande2023_tachyon}
A.~Ande, V.~Subhash, and V.~Natarajan.
\newblock {TACHYON: Efficient Shared Memory Parallel Computation of Extremum Graphs}.
\newblock {\em Computer Graphics Forum}, 42(6):e14784, 2023. doi: {{%
10\hspace{.1pt}\discretionary{.}{%
}{.}\hspace{.4pt}1111\discretionary{/}{%
}{/}cgf\hspace{.1pt}\discretionary{.}{%
}{.}\hspace{.4pt}14784}}


\bibitem{Ayachit2015ParaView}
U.~Ayachit.
\newblock {\em The {ParaView} guide, a parallel visualization application}.
\newblock Kitware, 2015.

\bibitem{Cai2017}
L.~Cai, B.~P. Nguyen, C.-K. Chui, and S.-H. Ong.
\newblock A two-level clustering approach for multidimensional transfer function specification in volume visualization.
\newblock {\em The Visual Computer}, 33:163--177, 2017.

\bibitem{Correa2011_TopoSpines}
C.~Correa, P.~Lindstrom, and P.-T. Bremer.
\newblock Topological spines: A structure-preserving visual representation of scalar fields.
\newblock {\em IEEE Transactions on Visualization and Computer Graphics}, 17(12):1842--1851, 2011. doi: {{%
10\hspace{.1pt}\discretionary{.}{%
}{.}\hspace{.4pt}1109\discretionary{/}{%
}{/}TVCG\hspace{.1pt}\discretionary{.}{%
}{.}\hspace{.4pt}2011\hspace{.1pt}\discretionary{.}{%
}{.}\hspace{.4pt}244}}


\bibitem{Dobrev2011}
P.~Dobrev, T.~van Long, and L.~Linsen.
\newblock A cluster hierarchy-based volume rendering approach for interactive visual exploration of multi-variate volume data.
\newblock In {\em Vision, Modeling, and Visualization (VMV'11)}, pp. 137--144, 2011.

\bibitem{edelsbrunner2002topological}
Edelsbrunner, Letscher, and Zomorodian.
\newblock Topological persistence and simplification.
\newblock {\em Discrete \& Computational Geometry}, 28:511--533, 2002.

\bibitem{edelsbrunner_simulation_1990}
H.~Edelsbrunner and E.~P. Mücke.
\newblock Simulation of simplicity: a technique to cope with degenerate cases in geometric algorithms.
\newblock {\em ACM Transactions of Graphics}, 9(1):66--104, Jan. 1990. doi: {{%
10\hspace{.1pt}\discretionary{.}{%
}{.}\hspace{.4pt}1145\discretionary{/}{%
}{/}77635\hspace{.1pt}\discretionary{.}{%
}{.}\hspace{.4pt}77639}}


\bibitem{engelke2021topology}
W.~Engelke, T.~B. Masood, J.~Beran, R.~Caballero, and I.~Hotz.
\newblock Topology-based feature design and tracking for multi-center cyclones.
\newblock In {\em Topological Methods in Data Analysis and Visualization VI: Theory, Applications, and Software}, pp. 71--85. Springer, 2021.

\bibitem{Falk2017}
M.~Falk, I.~Hotz, P.~Ljung, D.~Treanor, A.~Ynnerman, and C.~Lundstr\"om.
\newblock Transfer function design toolbox for full-color volume datasets.
\newblock In {\em {IEEE} PacificVis '17}, 2017.

\bibitem{Flohr:2023:PCCD}
T.~Flohr, B.~Schmidt, S.~Ulzheimer, and H.~Alkadhi.
\newblock Cardiac imaging with photon counting ct.
\newblock {\em British Journal of Radiology}, 96(1152):20230407, 10 2023. doi: {{%
10\hspace{.1pt}\discretionary{.}{%
}{.}\hspace{.4pt}1259\discretionary{/}{%
}{/}bjr\hspace{.1pt}\discretionary{.}{%
}{.}\hspace{.4pt}20230407}}


\bibitem{forman1998morse}
R.~Forman.
\newblock Morse theory for cell complexes.
\newblock {\em Advances in mathematics}, 134(1):90--145, 1998.

\bibitem{Greffier2025}
J.~Greffier, A.~Viry, A.~Rober, M.~Khorsi, and S.~Si-Mohamed.
\newblock Photon-counting ct systems: A technical review of current clinical possibilities.
\newblock {\em Diagnostic and Interventional Imaging}, 106:53--59, 2025.

\bibitem{Gyulassy2016_ion_battery}
A.~Gyulassy, A.~Knoll, K.~C. Lau, B.~Wang, P.-T. Bremer, M.~E. Papka, L.~A. Curtiss, and V.~Pascucci.
\newblock Interstitial and interlayer ion diffusion geometry extraction in graphitic nanosphere battery materials.
\newblock {\em IEEE Transactions on Visualization and Computer Graphics}, 22(1):916--925, 2016. doi: {{%
10\hspace{.1pt}\discretionary{.}{%
}{.}\hspace{.4pt}1109\discretionary{/}{%
}{/}TVCG\hspace{.1pt}\discretionary{.}{%
}{.}\hspace{.4pt}2015\hspace{.1pt}\discretionary{.}{%
}{.}\hspace{.4pt}2467432}}


\bibitem{Haidacher2008}
M.~Haidacher, S.~Bruckner, A.~Kanitsar, and E.~Gr\"oller.
\newblock Information-based transfer functions for multimodal visualization.
\newblock {\em Eurographics Workshop on Visual Computing for Biomedicine (VCBM)}, 2008.

\bibitem{HeineScalar2016}
C.~Heine, H.~Leitte, M.~Hlawitschka, F.~Iuricich, L.~De~Floriani, G.~Scheuermann, H.~Hagen, and C.~Garth.
\newblock A survey of topology-based methods in visualization.
\newblock {\em Computer Graphics Forum}, 35(3):643--667, 2016. doi: {{%
10\hspace{.1pt}\discretionary{.}{%
}{.}\hspace{.4pt}1111\discretionary{/}{%
}{/}cgf\hspace{.1pt}\discretionary{.}{%
}{.}\hspace{.4pt}12933}}


\bibitem{Huang2025_bimodal}
X.~Huang, H.~Miao, H.~Kim, A.~Townsend, K.~Champley, J.~Tringe, V.~Pascucci, and P.-T. Bremer.
\newblock Bimodal visualization of industrial x-ray and neutron computed tomography data.
\newblock {\em IEEE Transactions on Visualization and Computer Graphics}, 31(4):2196--2210, 2025. doi: {{%
10\hspace{.1pt}\discretionary{.}{%
}{.}\hspace{.4pt}1109\discretionary{/}{%
}{/}TVCG\hspace{.1pt}\discretionary{.}{%
}{.}\hspace{.4pt}2024\hspace{.1pt}\discretionary{.}{%
}{.}\hspace{.4pt}3382607}}


\bibitem{Jankowai2020}
J.~Jankowai, R.~Sk{\aa}nberg, D.~J{\"o}nsson, A.~Ynnerman, and I.~Hotz.
\newblock Tensor volume exploration using feature space representatives.
\newblock In {\em LEVIA 20}, 2020.

\bibitem{inviwo2019}
D.~J{\"o}nsson, P.~Steneteg, E.~Sund{\'e}n, R.~Englund, S.~Kottravel, M.~Falk, A.~Ynnerman, I.~Hotz, and T.~Ropinski.
\newblock {Inviwo} -- a visualization system with usage abstraction levels.
\newblock {\em IEEE Transactions on Visualization and Computer Graphics}, 26(11):3241--3254, 2019. doi: {{%
10\hspace{.1pt}\discretionary{.}{%
}{.}\hspace{.4pt}1109\discretionary{/}{%
}{/}TVCG\hspace{.1pt}\discretionary{.}{%
}{.}\hspace{.4pt}2019\hspace{.1pt}\discretionary{.}{%
}{.}\hspace{.4pt}2920639}}


\bibitem{Kim2011}
H.~S. Kim, J.~P. Schulze, A.~C. Cone, G.~E. Sosinsky, and M.~E. Martone.
\newblock Dimensionality reduction on multi-dimensional transfer functions for multi-channel volume data sets.
\newblock {\em Information Visualization}, 9(3):167--180, 2011. doi: {{%
10\hspace{.1pt}\discretionary{.}{%
}{.}\hspace{.4pt}1057\discretionary{/}{%
}{/}ivs\hspace{.1pt}\discretionary{.}{%
}{.}\hspace{.4pt}2010\hspace{.1pt}\discretionary{.}{%
}{.}\hspace{.4pt}6}}


\bibitem{Kniss2002}
J.~Kniss, G.~Kindlmann, and C.~Hansen.
\newblock {Multidimensional transfer functions for interactive volume rendering}.
\newblock {\em IEEE Transactions on Visualization and Computer Graphics}, 8(3):270--285, 2002. doi: {{%
10\hspace{.1pt}\discretionary{.}{%
}{.}\hspace{.4pt}1109\discretionary{/}{%
}{/}TVCG\hspace{.1pt}\discretionary{.}{%
}{.}\hspace{.4pt}2002\hspace{.1pt}\discretionary{.}{%
}{.}\hspace{.4pt}1021579}}


\bibitem{Konst2024PCCT}
B.~Konst, L.~Ohlsson, L.~Henriksson, M.~Sandstedt, A.~Persson, and T.~Ebbers.
\newblock Optimization of photon counting {CT} for cardiac imaging in patients with left ventricular assist devices: An in‐depth assessment of metal artifacts.
\newblock {\em Journal of Applied Clinical Medical Physics}, 25, 05 2024. doi: {{%
10\hspace{.1pt}\discretionary{.}{%
}{.}\hspace{.4pt}1002\discretionary{/}{%
}{/}acm2\hspace{.1pt}\discretionary{.}{%
}{.}\hspace{.4pt}14386}}


\bibitem{Liu2014}
S.~Liu, B.~Wang, J.~J. Thiagarajan, P.-T. Bremer, and V.~Pascucci.
\newblock Multivariate volume visualization through dynamic projections.
\newblock In {\em IEEE Symposium on Large Data Analysis and Visualization (LDAV)}, 2014. doi: {{%
10\hspace{.1pt}\discretionary{.}{%
}{.}\hspace{.4pt}1109\discretionary{/}{%
}{/}LDAV\hspace{.1pt}\discretionary{.}{%
}{.}\hspace{.4pt}2014\hspace{.1pt}\discretionary{.}{%
}{.}\hspace{.4pt}7013202}}


\bibitem{Ljung2016}
P.~Ljung, J.~Kr{\"u}ger, E.~Gr\"oller, M.~Hadwiger, C.~D. Hansen, and A.~Ynnerman.
\newblock State of the art in transfer functions for direct volume rendering.
\newblock {\em Computer Graphics Forum}, 35(3):23, 2016.

\bibitem{Ljung2016TF}
P.~Ljung, J.~Krüger, E.~Gröller, M.~Hadwiger, C.~D. Hansen, and A.~Ynnerman.
\newblock {State of the Art in Transfer Functions for Direct Volume Rendering}.
\newblock {\em Computer Graphics Forum}, 2016. doi: {{%
10\hspace{.1pt}\discretionary{.}{%
}{.}\hspace{.4pt}1111\discretionary{/}{%
}{/}cgf\hspace{.1pt}\discretionary{.}{%
}{.}\hspace{.4pt}12934}}


\bibitem{nilsson2022exploring}
E.~Nilsson, J.~Lukasczyk, W.~Engelke, T.~B. Masood, G.~Svensson, R.~Caballero, C.~Garth, and I.~Hotz.
\newblock Exploring cyclone evolution with hierarchical features.
\newblock In {\em Topological Data Analysis and Visualization (TopoInVis)}, pp. 92--102. IEEE, 2022.

\bibitem{Noordmans2000}
H.~Noordmans, H.~van~der Voort, and A.~Smeulders.
\newblock Spectral volume rendering.
\newblock {\em {IEEE Transactions on Visualization and Computer Graphics (TVCG)}}, 6(3):196--207, 2000. doi: {{%
doi\discretionary{:}{%
}{:}10\hspace{.1pt}\discretionary{.}{%
}{.}\hspace{.4pt}1109\discretionary{/}{%
}{/}2945\hspace{.1pt}\discretionary{.}{%
}{.}\hspace{.4pt}879782\hspace{.1pt}\discretionary{.}{%
}{.}\hspace{.4pt} 2}}


\bibitem{Pandey2021_Granular}
K.~Pandey, T.~Bin~Masood, S.~Singh, I.~Hotz, V.~Natarajan, and T.~G. Murthy.
\newblock Morse theory-based segmentation and fabric quantification of granular materials.
\newblock {\em Granular Matter}, 24(1):27, Dec 2021. doi: {{%
10\hspace{.1pt}\discretionary{.}{%
}{.}\hspace{.4pt}1007\discretionary{/}{%
}{/}s10035\discretionary{%
}{-}{-}021\discretionary{%
}{-}{-}01182\discretionary{%
}{-}{-}7}}


\bibitem{Shivashankar2016_Felix}
N.~Shivashankar, P.~Pranav, V.~Natarajan, R.~v.~d. Weygaert, E.~P. Bos, and S.~Rieder.
\newblock Felix: A topology based framework for visual exploration of cosmic filaments.
\newblock {\em IEEE Transactions on Visualization and Computer Graphics}, 22(6):1745--1759, 2016. doi: {{%
10\hspace{.1pt}\discretionary{.}{%
}{.}\hspace{.4pt}1109\discretionary{/}{%
}{/}TVCG\hspace{.1pt}\discretionary{.}{%
}{.}\hspace{.4pt}2015\hspace{.1pt}\discretionary{.}{%
}{.}\hspace{.4pt}2452919}}


\bibitem{sousbie2011persistent}
T.~Sousbie.
\newblock The persistent cosmic web and its filamentary structure--i. theory and implementation.
\newblock {\em Monthly Notices of the Royal Astronomical Society}, 414(1):350--383, 2011.

\bibitem{Strengert2006}
M.~Strengert, T.~Klein, R.~Botchen, S.~Stegmaier, M.~Chen, and T.~Ertl.
\newblock Spectral volume rendering using {GPU-based} raycasting.
\newblock {\em The Visual Computer}, 22(8):550--561, 2006. doi: {{%
10\hspace{.1pt}\discretionary{.}{%
}{.}\hspace{.4pt}1007\discretionary{/}{%
}{/}s00371\discretionary{%
}{-}{-}006\discretionary{%
}{-}{-}0028\discretionary{%
}{-}{-}0}}


\bibitem{Tierny2018ttk}
J.~Tierny, G.~Favelier, J.~A. Levine, C.~Gueunet, and M.~Michaux.
\newblock {The Topology ToolKit}.
\newblock {\em IEEE Transactions on Visualization and Computer Graphics}, 24(1):832--842, 2018. doi: {{%
10\hspace{.1pt}\discretionary{.}{%
}{.}\hspace{.4pt}1109\discretionary{/}{%
}{/}TVCG\hspace{.1pt}\discretionary{.}{%
}{.}\hspace{.4pt}2017\hspace{.1pt}\discretionary{.}{%
}{.}\hspace{.4pt}2743938}}


\bibitem{Bie2023}
J.~{van der Bie}, M.~{van Straten}, R.~Booij, D.~Bos, M.~L. Dijkshoorn, A.~Hirsch, S.~P. Sharma, E.~H. Oei, and R.~P. Budde.
\newblock Photon-counting {CT}: Review of initial clinical results.
\newblock {\em European Journal of Radiology}, 163:110829, 2023. doi: {{%
doi\hspace{.1pt}\discretionary{.}{%
}{.}\hspace{.4pt}org\discretionary{/}{%
}{/}10\hspace{.1pt}\discretionary{.}{%
}{.}\hspace{.4pt}1016\discretionary{/}{%
}{/}j\hspace{.1pt}\discretionary{.}{%
}{.}\hspace{.4pt}ejrad\hspace{.1pt}\discretionary{.}{%
}{.}\hspace{.4pt}2023\hspace{.1pt}\discretionary{.}{%
}{.}\hspace{.4pt}110829}}


\bibitem{Wang2012}
Y.~Wang, J.~Zhang, D.~J. Lehmann, H.~Theisel, and X.~Chi.
\newblock Automating transfer function design with valley cell-based clustering of {2D} density plots.
\newblock {\em Computer Graphics Forum}, 31(3), 2012.

\bibitem{Wu2019}
X.~Wu, P.~He, Z.~Long, X.~Guo, M.~Chen, X.~Ren, P.~Chen, L.~Deng, K.~An, P.~Li, B.~Wei, and P.~Feng.
\newblock Multi-material decomposition of spectral {CT} images via fully convolutional {DenseNets}.
\newblock {\em Journal of X-Ray Sciences and Technology}, 27:461--471, 2019. doi: {{%
DOI 10\hspace{.1pt}\discretionary{.}{%
}{.}\hspace{.4pt}3233\discretionary{/}{%
}{/}XST\discretionary{%
}{-}{-}190500}}


\bibitem{YanScalar2021}
L.~Yan, T.~B. Masood, R.~Sridharamurthy, F.~Rasheed, V.~Natarajan, I.~Hotz, and B.~Wang.
\newblock Scalar field comparison with topological descriptors: Properties and applications for scientific visualization.
\newblock {\em Computer Graphics Forum}, 40(3):599--633, 2021. doi: {{%
10\hspace{.1pt}\discretionary{.}{%
}{.}\hspace{.4pt}1111\discretionary{/}{%
}{/}cgf\hspace{.1pt}\discretionary{.}{%
}{.}\hspace{.4pt}14331}}


\bibitem{Zhao2010}
X.~Zhao and A.~E. Kaufman.
\newblock Multi-dimensional reduction and transfer function design using parallel coordinates.
\newblock In {\em IEEE/EG International conference on Volume Graphics}, pp. 69--76, 2010.

\end{thebibliography}
\end{document}